\documentclass[preprint,1p]{elsarticle}

\usepackage{amssymb}
\usepackage{graphicx}
\usepackage{subfigure}
\usepackage{pstricks-add}
\usepackage{pst-3dplot}	

\journal{Nuclear Physics B}

\begin{document}

\begin{frontmatter}

\title{Gravitational instabilities of isothermal spheres in the presence of a cosmological constant}

\author[demo]{Minos Axenides,}
\author[demo]{George Georgiou}
\author[demo,ntua]{Zacharias Roupas\corref{cor1}}
\ead{roupas@inp.demokritos.gr}

\address[demo]{Institute of Nuclear and Particle Physics, N.C.S.R. Demokritos, GR-15310 Athens, Greece}
\address[ntua]{Physics Department, National Technical University of Athens, GR-15780, Athens, Greece}

\cortext[cor1]{Corresponding author}

\begin{abstract}
Gravitational instabilities of isothermal spheres are studied in the presence of a positive or negative cosmological constant, in the Newtonian limit. In gravity, the statistical ensembles are not equivalent. We perform the analysis both in the microcanonical and the canonical ensembles, for which the corresponding instabilities are known as `gravothermal catastrophe' and `isothermal collapse', respectively. In the microcanonical ensemble, no equilibria can be found for radii larger than a critical value, which is increasing with increasing cosmological constant. In contrast, in the canonical ensemble, no equilibria can be found for radii smaller than a critical value, which is decreasing with increasing cosmological constant. For a positive cosmological constant, characteristic reentrant behavior is observed.	
\end{abstract}


\end{frontmatter}

\section{Introduction}
\indent In a seminal work \cite{Antonov}, Antonov described a thermodynamic instability of self-gravitating systems in the microcanonical ensemble, that later became known as `gravothermal catastrophe' \cite{Bell-Wood}. A classic review on thermodynamics and statistical mechanics of self-gravitating systems is the one of Padmanabhan \cite{Padman} and a more recent one is written by Katz \cite{Katz}. An extension of Antonov's instability to the canonical ensemble, named `isothermal collapse', was given by Chavanis \cite{Chavanis1}. Extended reviews on the self-gravitating gas at thermal equilibrium, from the statistical mechanics point of view, are given by de Vega \& Sanchez \cite{devegaSanchez1,devegaSanchez2} and Destri \& de Vega \cite{destridevega}. Thermodynamics of self-gravitating systems can be realized as the pioneering part of a, nowadays, more general, rapidly developing, new field of research, that is the thermodynamics of systems with long-range interactions \cite{Bell,LRI1,LRI2}. \\
\indent In a recent letter \cite{AGR} we reported on the effect of the cosmological constant to the Antonov's gravito-thermal instability in the microcanonical ensemble. In the present work not only do we enter in the details of this analysis and report some new results, but also extend the analysis to the canonical ensemble \cite{AGR2}. The two ensembles provide qualitatively different results. The understanding of the cosmological constant is of great importance mainly due to dark energy on the one hand (positive cosmological constant) and AdS/CFT correspondence on the other (negative cosmological constant). For convenience we shall call a positive cosmological constant `dS case', a negative one `AdS case' and a zero one `flat case', although we are working in the Newtonian limit, for which the de Sitter and anti-de Sitter spaces are more properly called `Newton-Hooke' spaces \cite{NewHook}. \\
\indent The original system \cite{Antonov,Bell-Wood} under study is a spherically bounded self-gravitating gas in the microcanonical ensemble, i.e. the spherical boundary shell has insulating and perfectly reflecting walls. The system is studied in the Newtonian limit and in the mean field approximation. Antonov proved that there is no global entropy maximum. Local entropy extrema (metastable states) exist only for $ER > -0.335GM^2$ and these equilibria are stable (entropy maxima) only if $\rho_0/\rho_R < 709$, where $E$, $R$ are the energy of the system and the radius of the shell and $\rho_0$, $\rho_R$ are the density of the centre and the edge, respectively. Lynden-Bell and Wood \cite{Bell-Wood} conjectured that at the region of no equilibrium, that is for $ER < -0.335GM^2$, the system would overheat and collapse. This gravothermal catastrophe picture was later confirmed by numerical simulations \cite{Sim1,Sim2,Sim3,devegaSanchez1,Sim4} and has been known as `core collapse' \cite{binney}, which plays a crucial role in the evolution of globular clusters. As indicated by de Vega and Sanchez \cite{devegaSanchez1} the collapse is a zeroth order phase transition, since the temperature and pressure increase discontinuously at the transition (the Gibbs free energy becomes discontinuous). Gravothermal catastrophe can also lead to the formation of supermassive black holes \cite{Shapiro}. \\
\indent The canonical ensemble of the system is studied by Chavanis \cite{Chavanis1}. From the statistical mechanics point of view, the canonical ensemble in gravity cannot be properly defined as explained by Padmanabhan \cite{Padman}. However, it can be defined formally by the use of free energy and can have physical realizations, as suggested in Refs. \cite{devegaNature,devegaApJ}. In Ref. \cite{devegaNature}, the interstellar medium is studied as a self-gravitating gas in thermal equilibrium with the microwave background. It is shown that self-gravity in the canonical ensemble can explain the fractal structure of interstellar medium. The very same mechanism is applied in \cite{devegaApJ} to explain the fractal structure of the Universe, i.e. the galaxy distributions, assuming galaxies have reached quasi-equilibrium. Chavanis \cite{Chavanis1} studied bounded isothermal spheres in the canonical ensemble and found that the self-similar behavior studied in Refs. \cite{devegaNature,devegaApJ,Semelin} originates in the secondary instabilities of bounded isothermal spheres that lead to a fragmented collapse, associated with the King's radius of the system. In contrast, the Jeans radius is associated with the isothermal non-fragmented collapse that occurs for $GM\beta/R > 2.52$. Our study in the canonical ensemble can be considered as a generalization of Chavanis' \cite{Chavanis1} study in the presence of a cosmological constant. \\
\indent The growing interest on Anti-de Sitter space, due to the AdS/CFT correspondence and the effect of the present value of the cosmological constant on the large scale structure of the Universe justify a stability analysis of gravitating systems in the presence of a cosmological constant term. In addition, in most modern cosmological models, such a term accompanies, the one or the other way, the evolution of the Universe from its beginning to the present. In cosmological models with a decaying vacuum energy \cite{Waga,Woodard,Polyakov} the cosmological constant is decreasing, so that its effect could be important even for stellar objects in the far past \cite{Boehmer}. For all these reasons, we believe it is crucial to understand the effect of (an arbitrary value of) the cosmological constant on the stability of gravitational systems. In this perspective, we study the simpler possible setup, i.e. a spherically symmetric, Newtonian, bounded system, to gain a basic understanding of the effect of the cosmological constant to the stability of self-gravitating systems. \\
\indent It has been proved by de Vega and Siebert \cite{devega,devega2} that a thermodynamic limit, different from the usual one, does exist for a self-gravitating gas in the presence of a cosmological constant and that the mean field approximation correctly describes this limit. In the presence of the cosmological constant we find that, in the mean field approximation and the Newtonian limit, the negative cosmological constant (AdS case) tends to destabilize the system, while the positive cosmological constant (dS case) tends to stabilize it. This result further supports recent investigations of AdS instabilities \cite{AdS1,AdS2}. In dS case many novel features arise. The system presents a reentrant behavior \cite{AGR}. A second critical radius, above which metastable states are restored, emerges, and in the canonical ensemble the system undergoes \textit{reentrant phase transition}, since there appear two critical temperatures. Reentrant phase transitions were known to occur for statistical systems with long-range interactions \cite{reentrant1,reentrant2,reentrant3,reentrant4} but not for gravitating systems. In addition, the homogeneous solution of dS has a turning point of stability, which we calculate analytically, and there exist infinite non-uniform solutions in the homogeneous radius as well as \textit{multiple series of equilibria} for any radius. \\
\indent The paper is organized as follows. In section \ref{sec:entropy} we calculate the entropy extrema, in section \ref{sec:criteria} we present the criteria for stability that we used in our analysis, in section \ref{sec:TE} the temperature and energy of the system are calculated and in section \ref{sec:series} the way to numerically generate the series of equilibria  is presented and their asymptotic behavior is analytically studied. In section \ref{sec:homo} is studied the stability of the homogeneous solution in dS in both ensembles. The main results of the microcanonical ensemble are presented in section \ref{sec:microcanonical}, where the correspondence of our `dS case' with the Schwartzschild-dS space is discussed, as well. The results in the canonical ensemble are given in section \ref{sec:canonical}.

\section{Entropy extremization and free energy}\label{sec:entropy}

Consider $N>>1$ identical particles (stars), bounded inside a spherical shell with insulating and perfectly reflecting walls. In order to calculate the entropy of the system, one should calculate the $N$-body distribution function
$f_N(\vec{r}_1,\ldots,\vec{r}_N,\vec{p}_1,\ldots,\vec{p}_N)$. This seems an impossible task. However, if the correlations between the particles are not significant and an intermediate scale where the granularity of the system can be ignored exists, one can work in the mean field approximation\cite{Padman} using the 1-body distribution function $f(\vec{r},\vec{p},t)$. This can be defined by the $N$-body distribution function as
\begin{equation}
	f(\vec{r}_1,\vec{p}_1,t) = \int{f_N(\vec{r}_1,\ldots,\vec{r}_N,\vec{p}_1,\ldots,\vec{p}_N)d^3\vec{r}_2\ldots d^3\vec{r}_Nd^3\vec{p}_2\ldots d^3\vec{p}_N}
\end{equation}
or one can think as $f$ giving the mass $dm$ inside a volume $d^3\vec{r}d^3\vec{p}$:
\begin{equation}
	dm = f(\vec{r},\vec{p},t)d^3\vec{r}d^3\vec{p}
\end{equation}
We assume that all particles have mass $\tilde{m} = 1$, so that we work with the velocity $\vec{\upsilon}$ instead of the momentum $\vec{p}$. Once $f$ has been determined, the density $\rho(\vec{r})$ can be found by
\begin{equation}\label{eq:rho}
	\rho(\vec{r}) = \int{f}d^3\vec{\upsilon}
\end{equation}
and the number of particles by
\begin{equation}\label{eq:N}
	N = \int{f} d^3\vec{r} d^3\vec{\upsilon}
\end{equation}
The total mass is $M = N \tilde{m}$.\\
The question posed is which $f$ extremizes the Boltzmann entropy
\begin{equation}\label{eq:entropy}
	S/k = - \int{f \log f d^6\tau }
\end{equation}
with constant energy $E$ and constant number of particles $N$, where $d^6\tau = d^3\vec{r} d^3\vec{\upsilon}$. For the case without a cosmological constant, it has been proved\cite{Antonov,Bell-Wood} that only spherical configurations maximize the entropy. For a discussion on spherical configurations in the presence of a cosmological constant see Ref.\cite{vir_appr}. We will consider only spherical, static distributions. In the derivation of $f$ we follow \cite{Bell-Wood}. \\
Using the Lagrange's multipliers $\beta = 1/kT$, $\mu$ the variation condition with respect to $f$ is
\begin{equation}\label{eq:Svariat}
	\delta S/k - \beta \delta E + \mu \delta N = 0
\end{equation}
The energy is $E = K + U$, where $K$ is the kinetic energy
\begin{equation}\label{eq:Kinetic}
	K = \frac{1}{2}\int{\upsilon^2f d^6\tau}
\end{equation}
Regarding the gravitational potential energy $U$, we have to work on the gravitational potential $\phi(\vec{r})$.
In the Newtonian limit (see \cite{Axenides} for dynamical effects of the cosmological constant in the Newtonian limit), the Poisson equation in the presence of a cosmological constant $\Lambda$ \cite{Wald} is
\begin{equation}\label{eq:poissonL}
	\nabla^2\phi = 4\pi G\rho - 8\pi G \rho_\Lambda 
\end{equation}
where $\rho_\Lambda = \frac{\Lambda c^2}{8\pi G}$. For an analytical derivation see Appendix \ref{app:B}.
The validity of the Newtonian approximation is analytically discussed in \cite{AGR}. In the Newtonian limit it should hold
\[
	\Lambda R^2 \ll 1
\]
while the cosmological constant is negligible if 
\[
	\rho \gg \rho_\Lambda \Leftrightarrow \Lambda R^2\frac{R}{R_S} \ll 1
\]
where $R_S = 2GM/c^2$ is the Schwartzschild radius of the system. Since for Newtonian systems it normally is $R/R_S \gg 1$, the cosmological constant is not in principle negligible. \\ 
\indent For spherically symmetric configurations, bounded in $r \in [0,R]$, the potential can be written as
\begin{equation}\label{eq:phi}
	\phi = \phi_N + \phi_\Lambda
\end{equation}
with
\begin{equation}\label{eq:phiN}
	\phi_N = -G\int_0^R{\frac{\rho(\vec{r}\,')}{|\vec{r}-\vec{r}\,'|}d^3\vec{r}\,'}
\end{equation}
\begin{equation}\label{eq:phiL}
	\phi_\Lambda = -\frac{4\pi G}{3}\rho_\Lambda r^2
\end{equation}
Therefore the potential energy can be written as
\begin{equation}\label{eq:U1}
	U = - \frac{G}{2} \int{\int{\frac{f(\vec{r} , \vec{\upsilon})f(\vec{r}\,' , \vec{\upsilon}\,')}{|\vec{r} - \vec{r}\,'|} d^6\tau d^6\tau '}} - \frac{4\pi G}{3}\rho_\Lambda \int{f r^2 d^6\tau}
\end{equation}
Using equations (\ref{eq:Kinetic}), (\ref{eq:U1}) the variation of energy is
\begin{eqnarray}\label{eq:Venergy}
	\delta E &=& \int{\delta f \frac{1}{2}\upsilon^2f d^6\tau} - \frac{G}{2} \int{\int{\frac{f'\delta f + f \delta f'}{|\vec{r} - \vec{r}\,'|} d^6\tau d^6\tau '}} - \frac{4\pi G}{3}\rho_\Lambda \int{\delta f r^2 d^6\tau} \nonumber\\
	&=& \int{\delta f \left( \frac{1}{2}\upsilon^2 - G \int{\frac{f'}{|\vec{r} - \vec{r}\,'|} d^6\tau '} - \frac{4\pi G}{3}\rho_\Lambda r^2 \right) d^6\tau} \nonumber \\
	&=& \int{\delta f \left( \frac{1}{2}\upsilon^2 + \phi \right)}
\end{eqnarray}
Using equations (\ref{eq:N}), (\ref{eq:entropy}) and (\ref{eq:Venergy}) we get
\begin{eqnarray}
	\delta S/k - \beta \delta E + \mu \delta N = -\int{\delta f \left(\log f + 1 + \beta \left(\frac{\upsilon^2}{2}+\phi\right) - \mu \right)d^6\tau}
\end{eqnarray}
So that, in order for equation (\ref{eq:Svariat}) to hold for all $\delta f$, we get
\begin{equation}\label{eq:distr1}
	\log f + 1 + \beta \left(\frac{\upsilon^2}{2}+\phi\right) - \mu = 0 \Rightarrow
	f(r,\upsilon) = A e^{-\beta\left(\frac{\upsilon^2}{2}+\phi(r)\right)}
\end{equation}
for $A = e^{\mu - 1}$. We get the Maxwell-Boltzmann distribution likewise `flat case'. The cosmological constant enters to the equation implicitly through the potential. \\
Now, the density distribution can easily be calculated
\begin{equation}\label{eq:density1}
	\rho = \int{f d^3\vec{\upsilon}} = \int{Ae^{-\frac{\beta \upsilon^2}{2}}e^{-\beta\phi}d^3\vec{\upsilon}} = 
	A\left(\frac{2\pi}{\beta}\right)^\frac{2}{3}e^{-\beta\phi}
\end{equation}
Absorbing the constants to the initial value $\phi(0)$ of the field and the central density $\rho_0$, we finally get
\begin{equation}\label{eq:density}
	\rho(r) = \rho_0 e^{-\beta(\phi(r)-\phi(0))}
\end{equation}
In spherical coordinates, the equation (\ref{eq:poissonL}), substituting equation (\ref{eq:density}), gives 
\begin{equation}\label{eq:poissonLsp}
	\frac{1}{r^2}\frac{d}{dr}\left( r^2\frac{d}{dr}\phi(r)\right) = 4\pi G\rho_0 e^{-\beta(\phi(r)-\phi(0))} - 8\pi G \rho_\Lambda 
\end{equation}
This equation holds for $r \leq R$, where $R$ is the radius of the bounding wall. For $r>R$, it is of course $\frac{1}{r^2}\frac{d}{dr}\left( r^2\frac{d}{dr}\phi(r)\right) = 0$ and $\phi$, $\phi'$ should be continuous at $R$. Let introduce the dimensionless variables
\begin{equation}\label{eq:variables}
	x = r\sqrt{4\pi G \rho_0 \beta} \; ,\; y = \beta (\phi - \phi(0)) \; , \; \lambda = \frac{2\rho_\Lambda}{\rho_0}
\end{equation}
Then, equation (\ref{eq:poissonLsp}) becomes
\begin{equation}\label{eq:emdenL}
	\frac{1}{x^2}\frac{d}{dx}\left( x^2\frac{d}{dx}y\right) = e^{-y} - \lambda
\end{equation}
which we call the Emden-$\Lambda$ equation. The initial conditions are
\begin{equation}\label{eq:cond}
	y(0) = 0 \; , \; y'(0) = 0
\end{equation}
The first initial condition comes from equation (\ref{eq:variables}), while the second from spherical symmetry (gravitational field at the center is zero). We call $z$ the value of $x$ at $R$:
\begin{equation}\label{eq:z}
	z = R\sqrt{4\pi G \rho_0 \beta}
\end{equation}
We want to generate series of equilibria, that is to find $y(z)$ for various $z$ , $\lambda$ (various isothermal spheres) and not just find $y(x)$ for some $\lambda$. This should be done by solving (\ref{eq:emdenL}) for various $z$, $\lambda$ keeping $N$, i.e. $M$, constant at each case. For the `flat' $\lambda = 0$ case this was very easy to perform. One could just solve Lane-Emden equation for one $z$ and interpret the result $y(x)$ as $y(z)$ with no inconsistency, since the mass could be considered fixed: there is no mass scale for this system. Another difficulty that enters, is that varying $\lambda$ cannot be realized as varying $\rho_\Lambda$, since $\lambda$ contains $\rho_0$, as well, that is different for each equilibrium configuration. Therefore, the situation in general becomes rather complex. We will see in section \ref{sec:TE}, how we resolved the problems by introducing a new parameter and constructing an appropriate computer code. \\
\indent The canonical ensemble can be studied by defining the Helmholtz free energy $F = E - TS$ with use of the Boltzmann entropy (\ref{eq:entropy}). It is equivalent to work with the Massieu function \cite{Katz1,Chavanis1} $J = -F/T$ that is 
\begin{equation}\label{eq:masieu}
	J = S-\frac{1}{T}E
\end{equation}
It is shown by Chavanis \cite{Chavanis1} that the maximization of $J$ with constant $T$ is equivalent to the maximization of $S$ with constant $E$ to first order in variations $\delta\rho$. This means that the two ensembles have the same equilibria defined by the distribution function 
\begin{equation}\label{eq:distr3}
	f = \left(\frac{\beta}{2\pi}\right)^{\frac{3}{2}}\rho_0 e^{-\beta(\phi - \phi(0))}e^{-\frac{1}{2}\beta\upsilon^2}
\end{equation} 
It is easy to check, performing for $J$ the previous calculations for $S$, that this holds true in the presence of a cosmological constant, too. What is different in the two ensembles is the stability analysis, i.e. the second order variation of entropy and free energy as we will see in section \ref{sec:criteria}.

\section{Criteria for stability}\label{sec:criteria}

We want to calculate the second order variation of entropy w.r.t. perturbations $\delta \rho$. We follow closely Padmanabhan \cite{Padman}. Maximizing the entropy for a given distribution $\rho(r)$ we get the Maxwell-Boltzmann distribution
\begin{equation}\label{eq:distr2}
	f(r,\upsilon) = \frac{1}{(2\pi kT)^{\frac{3}{2}}}\rho (r) e^{-\frac{\upsilon^2}{2kT}}
\end{equation}
The entropy can therefore be written as
\[
S/k = \frac{3M}{2}\log T - \int{\rho\log \rho d^3\vec{r}}
\]
We vary $\rho(r)$ and therefore $U$, $K$ through $\phi$, $T$ respectively keeping $E$ and $M$ fixed. Keeping $E$ fixed gives the constraint
\[
	\delta K + \delta U = 0
\]
From this equation we derive in \ref{app:C} that
\begin{equation}\label{eq:dT}
	\delta T = -\frac{2}{3M}\int d^3\vec{r} (\phi\delta\rho + \frac{1}{2}\delta\rho\delta\phi) + \mathcal{O}(3)
\end{equation}
So that, we get
\begin{eqnarray}
	\delta S/k + \mu \delta M &=& \frac{3M}{2T}\left(-\frac{2}{3M}\right)\int{d^3\vec{r}(\phi\delta\rho + \frac{1}{2}\delta\rho\delta\phi)} - \int{d^3\vec{r}\,\delta\rho(1+\log\rho)}  \nonumber \\
	&&\; - \frac{3M}{4T^2}
	\left( -\frac{2}{3M}\int{d^3\vec{r}(\phi\delta\rho+\frac{1}{2}\delta\rho\delta\phi)}\right)^2
	-\int{d^3\vec{r}\frac{(\delta\rho)^2}{2\rho}} \nonumber \\
	&&\; + \mu\int d^3\vec{r}\delta\rho + \mathcal{O}(3)
	= -\int{d^3\vec{r}\,\delta\rho(1+log\rho-\mu+\frac{\phi}{T})} \nonumber \\
	&&\; - \int{d^3\vec{r}\left( \frac{\delta\rho\delta\phi}{2T} + \frac{(\delta\rho)^2}{2\rho}\right) }
	 - \frac{1}{3MT^2}\left( \int{d^3\vec{r}\,\phi\delta\rho} \right)^2 + \mathcal{O}(3) \nonumber
\end{eqnarray}
with the second order variation being
\begin{equation}\label{eq:dS2}
	\delta^{(2)}S/k = - \int{d^3\vec{r}\left( \frac{\delta\rho\delta\phi}{2T} + \frac{(\delta\rho)^2}{2\rho}\right) }
	 - \frac{1}{3MT^2}\left( \int{d^3\vec{r}\,\phi\delta\rho} \right)^2
\end{equation}
If at an equilibrium, for any perturbation $\delta\rho$ it is $\delta^{(2)}S|_{equil} < 0$ then the entropy is a (local) maximum and the equilibrium is (locally) stable. If there exists one or more perturbations for which $\delta^{(2)}S|_{equil} > 0$, then the equilibrium is unstable. Let us see how the sign of $\delta^{(2)}S|_{equil}$ can be deduced by an eigenvalue equation \cite{Padman} generated by (\ref{eq:dS2}).  \\
Since the total mass is constant, for $\delta \rho$ should hold:
\begin{equation}\label{eq:drho}
	\int_0^R d^3 \vec{x} \delta\rho = 0
\end{equation}
\indent Let us concentrate on spherical symmetric perturbation $\delta \rho = \delta\rho(r)$ and introduce the mass perturbation
\begin{equation}\label{eq:qdef}
	q(r) = \delta M(r)
\end{equation}
Then
\begin{equation}\label{eq:qdrho}
	\delta \rho = \frac{1}{4\pi r^2} \frac{dq}{dr}
\end{equation}
The force due to perturbed distribution $(\delta\phi)' = G\frac{q}{r^2}$ has to be finite everywhere and therefore $q$ should go like $q\rightarrow r^3$ for $r\rightarrow 0$. This means that $q(0)=0$. Then equation (\ref{eq:drho}) gives that $q(R)=0$. Thus, the boundary conditions are
\begin{equation}\label{eq:qbound}
	q(0) = q(R) = 0
\end{equation}
Substituting (\ref{eq:qdef}), (\ref{eq:qdrho}) into equation (\ref{eq:dS2}) and performing several integrations by part, we get
\begin{eqnarray}\label{eq:d2Sbeta}
	\delta^{(2)}S &=&  - \frac{1}{3MT^2}\left( \int_0^R{dr \phi q'} \right)^2
	- \int_0^R{dr \left( \frac{ q'\delta\phi}{2T} + \frac{(q')^2}{8\pi\rho r^2}\right) } \nonumber \\
	&=& - \frac{1}{3MT^2}\left( \int_0^R{dr \phi ' q} \right)^2
	+ \frac{1}{2T} \int_0^R{dr q(\delta\phi)'} + \int_0^R{dr q\frac{d}{dr}\left(\frac{q'}{8\pi\rho r^2}\right) } \nonumber\\
	&=&  - \frac{1}{3MT^2}\left( \int_0^R{dr \phi ' q} \right)^2
	+ \frac{1}{2} \int_0^R{dr q \left\lbrace \frac{G}{Tr^2} + \frac{d}{dr}\left(\frac{1}{4\pi\rho r^2}\frac{d}{dr}\right)\right\rbrace }q 
\end{eqnarray}
The last expression can be written as
\begin{equation}
		\delta^{(2)}S = \int_0^R\int_0^R dr dr' q(r')\hat{K}(r,r')q(r) 
\end{equation}		
with
\begin{equation}
		\hat{K} = - \frac{\phi'(r)\phi'(r')}{3MT^2} 
	+ \frac{1}{2} \delta (r-r') \left\lbrace \frac{G}{Tr^2} + \frac{d}{dr}\left(\frac{1}{4\pi\rho r^2}\frac{d}{dr}\right)\right\rbrace 
\end{equation}
The sign of $\delta^{(2)}S$ is therefore determined by the eigenvalues of the `matrix' $K(r,r')$
\begin{equation}\label{eq:eigen1}
	\int_0^Rdr' \hat{K}(r,r') F_\xi(r') = \xi F(r)
\end{equation}
At this equilibrium, where there is a transition from stability to instability, it should be $\xi = 0$. If for an equilibrium is found one perturbation $F_\xi$ for which $\xi > 0$ then this equilibrium is unstable. For an equilibrium to be (locally) stable all eigenvalues should be negative $\xi  < 0$ for all perturbations. Equation (\ref{eq:eigen1}) gives
\begin{eqnarray}\label{eq:eigen2}
	 &&  - \frac{\phi'(r)}{3MT^2}\int_0^R\phi'(r')F_\xi (r')
	+ \frac{1}{2} \left\lbrace \frac{G}{Tr^2} + \frac{d}{dr}\left(\frac{1}{4\pi\rho r^2}\frac{d}{dr}\right)\right\rbrace F_\xi(r) = \xi F_\xi(r) \nonumber \\
	&& \left\lbrace \frac{G}{Tr^2} + \frac{d}{dr}\left(\frac{1}{4\pi\rho r^2}\frac{d}{dr}\right) -2\xi \right\rbrace F_\xi(r) = \frac{2V}{3MT^2} \phi'
\end{eqnarray}
with
\begin{equation}\label{eq:V1}
	V = \int_0^Rdr' \phi'(r')F_\xi(r')
\end{equation}
In (\ref{eq:eigen2}) the cosmological constant enters implicitly, since:
\[
	\phi' = \frac{GM(r)}{r^2} - \frac{8\pi G}{3}\rho_\Lambda r
\]
 The boundary conditions of (\ref{eq:eigen2}) are as given in (\ref{eq:qbound})
\begin{equation}\label{eq:Fbound}
	F_\xi(0) = F_\xi(R) = 0
\end{equation}
\indent We developed an algorithm that can determine eigenvalues and eigenstates for the boundary value problem defined by equations (\ref{eq:eigen2}), (\ref{eq:V1}), (\ref{eq:Fbound}). The main difficulty is that in $V$ enters the unknown function $F_\xi$. We resolve the problem as follows. For a given range of $\xi$, the problem is solved for trial values of $V$, call them $V_T$, and then the integral (\ref{eq:V1}) is calculated, which gives some value $\tilde{V}$. Some value $\xi$ is indeed an eigenvalue, only if $\tilde{V} = V_T$ and, in this case, of course $V = \tilde{V}=V_T$. The algorithm is applied for the dimensionless version of (\ref{eq:eigen2}), namely:
\begin{equation}\label{eq:eigen3}
 \left\lbrace \frac{1}{x^2} + \frac{d}{dx}\left(\frac{e^y}{x^2}\frac{d}{dx}\right) -\bar{\xi} \right\rbrace F_{\bar{\xi}} = \frac{2y'}{3Bz} \bar{V} 
\end{equation}
where
\[
	B = \frac{GM\beta}{R}\; , \;
	\bar{\xi} = \frac{\xi}{2\pi G^2\beta^2\rho_0} \; , \;
	\bar{V} = \beta V = \beta \int_0^z dx\, y' F_{\bar{\xi}}
\]
and $z$, $y$, $y'$, $B$ are calculated at the equilibrium. Using the algorithm we can determine a turning point, where $\xi = 0$, an instability $\xi > 0$ or verify a stable branch of series of equilibria by checking every equilibrium point for a zero eigenvalue and for an as large as possible range of positive eigenvalues. We performed these tasks for every series of equilibria demonstrated in this paper. \\
\indent In an equilibrium for which $\xi = 0$, there is a transition from a stable branch to an unstable branch or from an unstable branch to a more unstable branch (or vice versa, of course). Suppose you approach the turning point from a stable series (all eigenvalues negative), then at the next equilibrium point after the transition, one eigenvalue becomes positive. In appendix \ref{app:D} we show how one can determine the branch with the additional positive eigenvalue (instability) near a turning point, from the previous analysis. \\
\indent Performing similar calculations for the second variation of free energy $J$, it is straightforward to find the corresponding eigenvalue problem for the canonical ensemble:
\begin{equation}\label{eq:canonEig}
	\left\lbrace\frac{G}{Tr^2} + \frac{d}{dr}\left(\frac{1}{4\pi\rho r^2}\frac{d}{dr}\right) -2\xi \right\rbrace F_\xi(r) = 0
\end{equation} 
Compared to eigenvalue equation (\ref{eq:eigen2}), we see that the difference is only the absence of the term containing the derivative of the potential. This difference changes the onset of the instability for the two ensembles. \\
\indent There is a way to study stability without solving an eigenvalue problem, due to a classical result of Poincar\'e \cite{Poincare}. In thermodynamics of self-gravitating systems, it was for the first time applied by Lynden-Bell \& Wood \cite{Bell-Wood}. Briefly, it states that a change of stability can only occur at a point, where two or more series of equilibria have a common point or where they merge into each other. As indicated by Katz \cite{Katz1}, practically this means, that in the microcanonical ensemble, the change of stability happens in the point where $\beta (E)$ has infinite slope, while in the canonical where $\beta (E)$ has extrema. Equivalently this means that $E$ in the microcanonical or $\beta$ in the canonical ensemble, has an extremum with respect to some other variable (e.g. the density contrast $\log(\rho_0/\rho_R)$ or $z$ in dimensionless variables) at the turning point. We determine the point of change of stability in the dimensionless variables by finding an extremum of $ER/GM^2$ in the microcanonical case and $GM\beta/R$ in the canonical case, with respect to the logarithm of the density contrast $\log\frac{\rho_0}{\rho_R}$. 	

\section{Temperature and Energy}\label{sec:TE}
At the thermodynamic equilibrium, the gas sphere has the same temperature $\beta$ everywhere, so that it is called an isothermal sphere. We define the dimensionless inverse temperature \cite{Bell-Wood}
\begin{equation}\label{eq:betabar}
	B = \frac{GM\beta}{R}
\end{equation}
which can be calculated with use of the dimensionless variables (\ref{eq:variables}), by integrating the Poisson with $\Lambda$ equation (\ref{eq:poissonLsp}):
\begin{eqnarray}\label{eq:Bdef}
	&& \left. R^2\frac{d\phi}{dr}\right|_R = GM - 8\pi G \rho_\Lambda \frac{R^3}{3} \Rightarrow 
	\frac{GM}{R} = \frac{z}{\beta} \left.\frac{dy}{dx}\right|_z + \frac{8\pi G}{3} \rho_\Lambda \frac{z^2}{4\pi G \rho_0 \beta}
	\Rightarrow \nonumber \\
	&& \frac{GM\beta}{R} \equiv B(z) = z y' + \frac{1}{3} \lambda z^2
\end{eqnarray}
The kinetic energy per particle is
\[
	\frac{K}{N} = \frac{\int{f\frac{\upsilon^2}{2}d^3\vec{\upsilon}}}{\int{f d^3\vec{\upsilon}}} = 
	\frac{\int{e^{-\frac{\upsilon^2}{2}}\frac{\upsilon^2}{2}d^3\vec{\upsilon}}}{\int{e^{-\frac{\upsilon^2}{2}}d^3\vec{\upsilon}}}
	= \frac{3}{2\beta}
\]
Since $M = N\tilde{m} = N$ we get
\begin{equation}\label{eq:Kenergy}
	K = \frac{3M}{2\beta}
\end{equation}
To calculate the total energy, we will use the Virial theorem, so that to avoid performing one more numerical integration to calculate the Newtonian potential energy and therefore improving the computer's performance. However, in the Appendix \ref{app:E} we perform a straightforward calculation of the energy in order to numerically cross-check the two expressions for some cases, who are proven to give identical results. This confirms the fact that bounded isothermal spheres are virialized in the presence of the cosmological constant. \\
The Virial theorem for a discrete collection of matter:
\[
	2<K> = - \sum_{k=1}^N<\vec{F}_k\cdot \vec{r}_k>
\]
can be generalized in our case as
\begin{equation}\label{eq:vir1}
	2 K = - \int{\rho (\nabla \phi \cdot \vec{r}) d^3\vec{r}} + 3PV
\end{equation}
where $V = \frac{4}{3}\pi R^3$ is the volume of the shell and $P$ the pressure exerted on the gas by the walls. The term $3PV$ arises simply since $F\cdot R = P\cdot 4\pi R^2\cdot R = 3PV$. The contribution of the Newtonian potential to the right-hand side is just the Newtonian potential energy $U_N$, while for the cosmological potential $\phi_\Lambda$ we get the term
\[
	\int{\rho (\nabla \phi_\Lambda \cdot \vec{r}) d^3\vec{r}} = \int{\rho \frac{8\pi G}{3}\rho_\Lambda r \cdot r d^3\vec{r}} = -2 U_\Lambda
\]
Substituting everything in equation (\ref{eq:vir1}) we get the following form of the Virial theorem:
\begin{equation}\label{eq:virial}
	2K + U_N - 2U_\Lambda = 3PV
\end{equation}
Eliminating $U_N$ we get
\begin{equation}\label{eq:virial}
	E = 3PV - K + 3U_\Lambda
\end{equation}
Using (\ref{eq:variables}) we write the potential $U_\Lambda$ in the dimensionless variables
\begin{eqnarray}\label{eq:UL}
	U_\Lambda &=&-\int{\rho  \frac{4\pi G}{3}\rho_\Lambda r^2 4\pi r^2 dr}
	  = -\frac{4\pi G}{3} \frac{2\rho_\Lambda}{\rho_0}\frac{\rho_0}{2} 
	 \int{\rho_0 e^{-y}  4\pi \frac{1}{(4\pi G \rho_0 \beta)^2} x^4 \frac{dx}{z/R}} \nonumber \\
	 &=& -\frac{\lambda}{6z}\frac{1}{G \beta^2 /R} \int{ e^{-y}   x^4 dx}
	  = -\frac{\lambda}{6z}\frac{1}{(G M\beta /R)^2} \frac{GM^2}{R}\int{ e^{-y}   x^4 dx} \nonumber \\
	 \frac{R}{GM^2}U_\Lambda &=& -\frac{\lambda}{6B^2 z}\int{ e^{-y}   x^4 dx}
\end{eqnarray}
Let us calculate the term $3PV$ in dimensionless variables
\[
	P = \frac{\rho}{\beta} = \frac{\rho_0 e^{-y}}{\beta} = \frac{z^2}{R^2 4\pi G \beta}\frac{e^{-y}}{\beta} = 
		\frac{z^2e^{-y}}{3 4\pi G/3}\frac{1}{(GM\beta/R)^2}\frac{GM^2}{R} \Rightarrow
		3PV\frac{R}{GM^2} = \frac{z^2e^{-y}}{B^2}
\]
We define the dimensionless energy
\begin{equation}\label{eq:Qenergy}
	Q = \frac{RE}{GM^2}
\end{equation}
We can calculate $Q$ from the virial equation (\ref{eq:virial})
\begin{equation}\label{eq:Eenergy}
	\frac{RE}{GM^2} \equiv Q(z) = \frac{z^2 e^{-y}}{B^2} - \frac{3}{2B} - \frac{\lambda}{2B^2 z}\int{ e^{-y}   x^4 dx}
\end{equation}
\begin{figure}[tb!]
\begin{center}
	\includegraphics[scale=0.8]{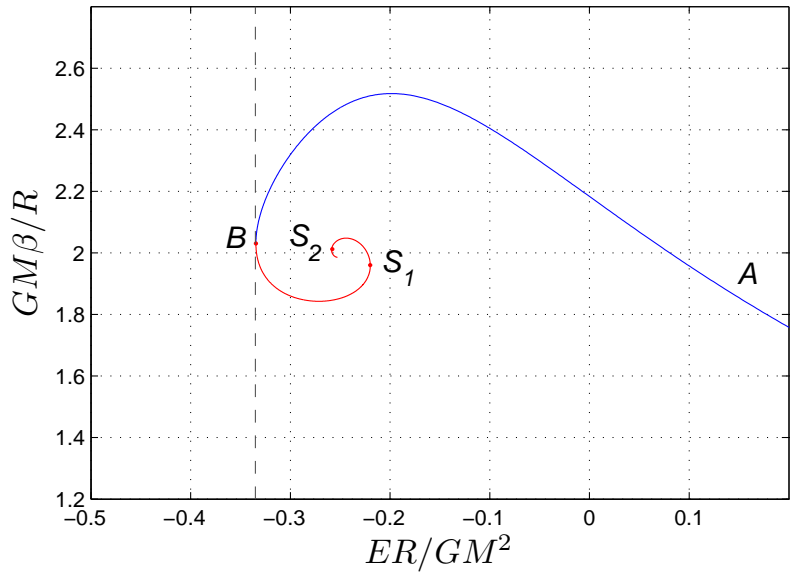} 
\end{center}
\caption{The series of equilibria $\beta = \beta(E)$ for $\Lambda = 0$. In the microcanonical ensemble, $AB$ is the stable branch and $B$ is the turning point of stability. As we can see, it is also the marginal point for which equilibrium states do exist.} 
\label{fig:spiral_flat}
\end{figure}
For simplicity let us focus in the microcanonical ensemble for the moment. The self-gravitating gas is characterized by two instabilities \cite{Padman}; a `strong' instability that is associated with complete absent of equilibria and a `weak' instability which refers to equilibria that are unstable, i.e. entropy, although it is an extremum, is not a local maximum. The equilibrium point at which the weak instability sets in, called the turning point, is therefore the point at which the second variation of entropy becomes zero, as calculated in the previous section. However, Poincar\'e's theorem insures \cite{Poincare,Katz1} that this point is the same with the marginal point of the strong instability. In Figure \ref{fig:spiral_flat}, where the series of equilibria $\beta = \beta(E)$ is drawn for the flat $\Lambda = 0$ case, this is point $B$. The strong instability corresponds to the region at the left of the vertical dashed line that crosses $B$. The weak instability refers to the branch $BS$, where $S$ is the focal point of the spiral. Thus, point $B$ is simultaneously the turning point of stability (from stable branch $AB$ to the unstable branch $BS$) and the marginal point of the strong instability. This strong instability leads to a core-halo structure as verified by monte-carlo simulations \cite{devegaSanchez1,Sim1,Sim2,Sim3,Sim4} and is associated with a collapse phase transition \cite{devegaSanchez1,devegaSanchez2,destridevega,devega2}, while the weak instability leads to a fractal structure and is associated with a fragmented collapse, called a clumping phase transition \cite{Chavanis1,devegaSanchez1,devegaSanchez2,destridevega,devega2}. This fractal structure is due to the secondary instabilities that set in at points $S_1$, $S_2$, etc. All of the above hold in the canonical ensemble, as well, where the axes in Figure \ref{fig:spiral_flat} should be interchanged and entropy should be replaced with free energy.

\section{The series of equilibria and asymptotic behavior}\label{sec:series}

\begin{figure}[here!]
\begin{center}
	\includegraphics[scale=0.8]{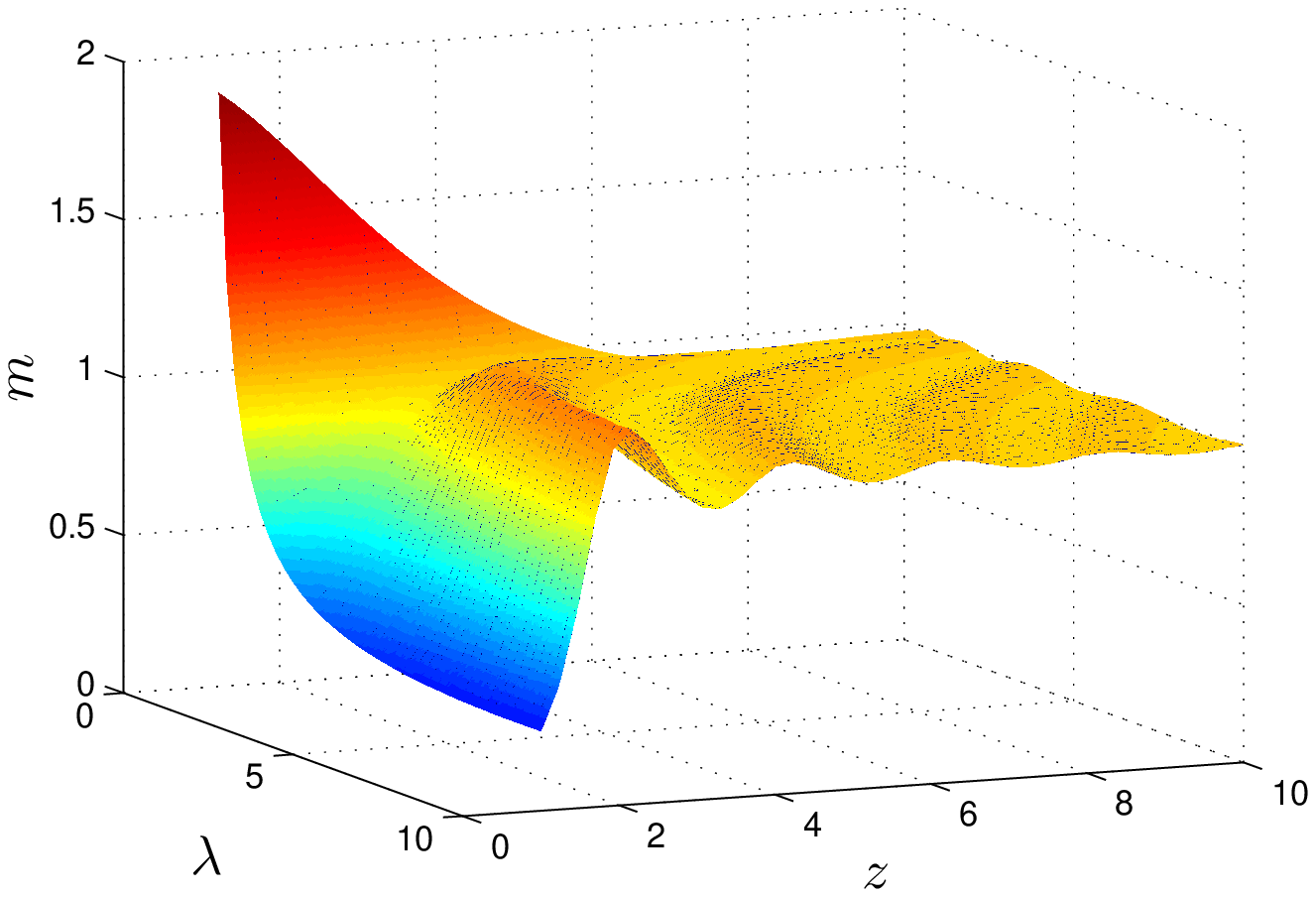} 
\end{center}
\caption{For $\rho_\Lambda >0$ (dS) there exist various series of equilibria for some fixed $\rho_\Lambda$.} 
\label{fig:m3D}
\end{figure}
\begin{figure}[tb]
\begin{center}
	\includegraphics[scale=0.8]{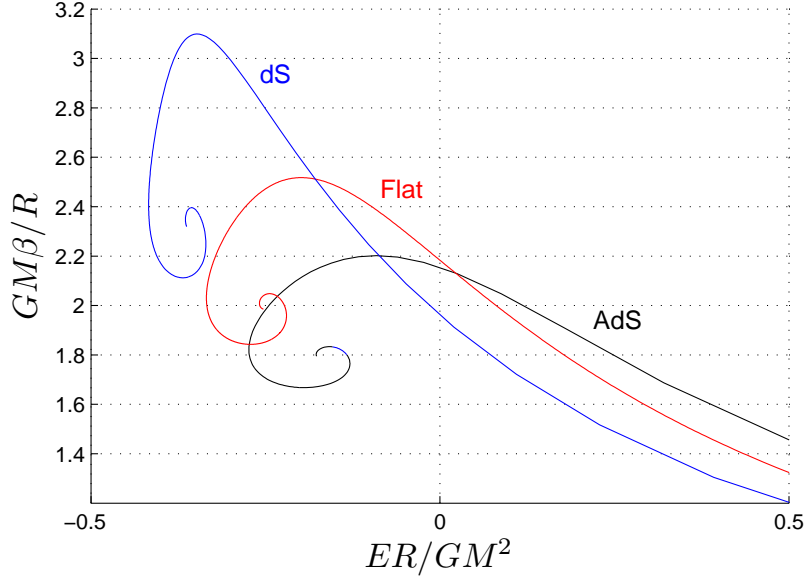} 
\end{center}
\caption{The temperature versus the energy for some negative (`AdS'), zero (`Flat') and positive (`dS') cosmological constant. The mass $M$ and the radius $R$ are held constant.} 
\label{fig:spirals}
\end{figure}
We want to solve the Emden-$\Lambda$ equation (\ref{eq:emdenL}) for various $\rho_\Lambda$ keeping $M$ constant and for various isothermal spheres with radius $R$. The cosmological constant introduces a mass scale
\[
	M_\Lambda = \frac{4}{3}\pi R^3 \cdot \rho_\Lambda
\]
We define the dimensionless mass
\begin{equation}\label{eq:mass1}
	m \equiv \frac{M}{2M_\Lambda} = \frac{\bar{\rho}}{2\rho_\Lambda}
\end{equation}
where $\bar{\rho} = M/(\frac{4}{3}\pi R^3)$ is the mean density of matter. This gives
\begin{equation}\label{eq:mass2}
	m = \frac{3}{8\pi}\frac{M}{R^3\rho_\Lambda}
\end{equation}
that is
\begin{equation}\label{eq:mass3} 
	m = \frac{\rho_0}{2\rho_\Lambda}\frac{1}{4\pi G \rho_0\beta R^2}\frac{3GM\beta}{R}\Rightarrow
	m = \frac{3B}{\lambda z^2}
\end{equation}
Equation (\ref{eq:mass3}) implies that in order to keep $m$ fixed, $\lambda$ has to be different at each $z$. We developed an algorithm that solves the Emden-$\Lambda$ equation keeping the quantity $m$ constant; for each $z$, some $\lambda$ values are iterated until the requested value of $m$, calculated by equation (\ref{eq:mass3}), is found for a relative tolerance predetermined by the user. For the various plots in this paper, we used relative tolerance between $10^{-7}-10^{-11}$ depending on the needs of each case. From equation (\ref{eq:mass2}), it is evident that solving for various fixed $m$ can be interpreted as varying $\rho_\Lambda$ and/or $R$ for a fixed $M$. Therefore we can determine how various quantities change with respect to $\rho_\Lambda$ by solving for various $m$. \\
\indent For the AdS case ($\rho_\Lambda < 0$) we find that for each $\rho_\Lambda$ only one series of equilibria exists, likewise flat ($\rho_\Lambda = 0$) case. However, for dS ($\rho_\Lambda > 0$), we find that for a fixed $\rho_\Lambda$, there exist more than one series of equilibria. This is evident in Figure \ref{fig:m3D} where $m$ is plotted w.r.t. $\lambda$, $z$. We see that the intersection of a plane $m=const$ with the $m$-surface defines various different curves in the $(\lambda,z)$ space. \\
\indent The Emden equation (\ref{eq:emdenL}) for $\rho_\lambda = 0$, i.e. $\lambda = 0$, is well known to have an exact but singular solution
\begin{equation}\label{eq:singFlat}
	y_s = \log\frac{z^2}{2}
\end{equation}
with infinite density at the origin since $e^{-y_s} = \frac{2}{z^2}$. It can be shown \cite{Bell-Wood} that for $z\rightarrow \infty$ the series of equilibria approach the singular solution. Therefore, the singular solution in the flat case is the focus point of the central spiral $\beta(E)$ in Figure \ref{fig:spirals}. We can see in this picture that in dS and AdS cases there seems to exist an equivalent to the flat singular solution, which can be identified with the focal points of the spirals (note that in dS, not all series of equilibria do form a spiral $\beta(E)$). Unfortunately, no analytical solution of equation (\ref{eq:emdenL}) is known. However, following the flat case paradigm we can determine asymptotically the equivalent of the singular solution in dS and AdS. Applying the transformations \cite{Bell-Wood}
\begin{equation}\label{eq:transfAsymp}
	\zeta = \log z \quad , \quad u = -y + 2\zeta
\end{equation}
the Emden-$\Lambda$ equation becomes
\begin{equation}\label{eq:EmdenTransf}
	\frac{d^2 u}{d\zeta^2} + \frac{du}{d\zeta} + e^u - 2 - \lambda e^{2\zeta} = 0
\end{equation}
This is the equation of an oscillator in the potential $V(u) = e^u - 2u$ with external force $F_{ext} = \lambda e^{2\zeta}- u'$
where prime denotes differentiation w.r.t. $\zeta$. The external force is damping in AdS case ($\lambda < 0$) and forced, damping in dS ($\lambda > 0$). The potential has a minimum $u_0 = \log 2$. In dS case, if the external force is damping, i.e. the term $-u'$ dominates, then for $\zeta \rightarrow \infty$, we have $u\rightarrow u_0$. Therefore, in this case, we can make an expansion of $u$ about $u_0$. The damping dominates if:
\begin{equation}\label{eq:asymptcondition}
	\lambda e^{2\zeta} \ll u' \Rightarrow \lambda z^2 \ll z y'\Rightarrow \lambda z^2 \ll \frac{m+2}{6}
\end{equation}
where $y'$ denotes differentiation w.r.t. $z$ and we have used equation (\ref{eq:mass3}). This limit means that we are considering very small $\rho_\Lambda$, just about the flat case, since $\lambda z^2 \rightarrow 0$ has to be taken together with $z \rightarrow \infty$. The two limits are consistent with each other, since by equation (\ref{eq:mass3}) we see that as $z \rightarrow \infty$ it must be $\lambda \rightarrow 0$ so as $\lambda z^2$ to remain finite in order for $m$ to be finite (and $B$ is bounded for the equilibria we are considering). \\
\indent Making the transformation
\[
	u = log 2+ u_1
\]
equation (\ref{eq:EmdenTransf}) becomes
\[
	\frac{d^2u_1}{d\zeta^2} + \frac{du_1}{d\zeta} + 2e^u_1 - 2 - \lambda e^{2\zeta} = 0
\]
Provided condition (\ref{eq:asymptcondition}) holds and for $\zeta \rightarrow\infty$, $u_1$ is small and we can expand the exponential keeping the first two terms, to get:
\[
	\frac{d^2 u_1}{d\zeta^2} + \frac{du_1}{d\zeta} + 2u_1 - \lambda e^{2\zeta} = 0
\]
The solution of the corresponding homogeneous equation is well known to be 
\[
u_h = Ae^{-\frac{\zeta}{2}}\cos(\frac{\sqrt{7}}{2}\zeta + \delta)
\]
and one solution of the full non-homogeneous equation can be found easily to be $u_p = \frac{\lambda}{8}e^{2\zeta}$. We get in total
\[
	u_1 = Ae^{-\frac{\zeta}{2}}\cos(\frac{\sqrt{7}}{2}\zeta + \delta) + \frac{\lambda}{8}e^{2\zeta}
\]
Since 
\[
	u = u_1 + \log2 \Rightarrow -y + 2\zeta = u_1  + \log2\Rightarrow y = \log \frac{z^2}{2} - u_1
\]
the asymptotic behavior of the Emden-$\Lambda$ equation for $z\rightarrow \infty$ and for small $\lambda z^2$ is
\begin{equation}\label{eq:asymptotic}
	y_a = \log \frac{z^2}{2} - \frac{A}{z^{\frac{1}{2}}}\cos(\frac{\sqrt{7}}{2}\log z + \delta) - \frac{\lambda}{8}z^2
\end{equation}
The density $\rho$ is given by the exponential $e^{-y}$. We get the asymptotic behavior
\begin{eqnarray}\label{eq:asymptotic2}
	e^{-y_a} &=& \frac{2}{z^2}e^{\frac{\lambda}{8}z^2}e^{A z^{-1/2}\cos(\frac{\sqrt{7}}{2}\log z + \delta)} \nonumber\\
	&\simeq& \frac{2}{z^2}e^{\frac{\lambda}{8}z^2}\left(1+\frac{A}{z^{\frac{1}{2}}}\cos(\frac{\sqrt{7}}{2}\log z + \delta)\right)
\end{eqnarray}
We see that the equivalence to the singular solution (\ref{eq:singFlat}) of the flat case, is given in dS and AdS by the `asymptotic singular' solution $e^{y_{AS}} = \frac{2}{z^2}e^{\frac{\lambda}{8}z^2}$ that is
\begin{equation}\label{eq:asymptSing}
	y_{AS} = \log \frac{z^2}{2} - \frac{\lambda}{8}z^2
\end{equation}
This solution $y_{AS}$, is easy to check that indeed satisfies the Emden-$\Lambda$ equation (\ref{eq:emdenL}) to first order in $\lambda z^2$. \\
By equations (\ref{eq:Bdef}) and (\ref{eq:Eenergy}) we can calculate the temperature and energy of the asymptotic singular solution to be:
\begin{equation}\label{eq:asymptsing_EB}
	\frac{GM\beta_{AS}}{R} = 2\frac{4m}{4m - 1}\quad , \quad \frac{RE_{AS}}{GM^2} = \frac{2e^{3B/8m}}{B^2} - \frac{3}{2B}-\frac{1}{mB} - \frac{9}{40m^2}
\end{equation}
Using this equation for $B_{AS}$, the condition (\ref{eq:asymptcondition}) gives the values of $m$ for which the asymptotic singular solution is valid
\begin{equation}\label{eq:asympt_mValid}
	m < -6.98 \quad \mbox{and}\quad m > 5.23
\end{equation}
\indent Recall that $m = \bar{\rho}/2\rho_\Lambda$ is used to vary the cosmological constant. So that $m\rightarrow\infty$ corresponds to the flat case, $m < 0$ to AdS and $m>0$ to dS. We find that the singular point determined analytically by equation (\ref{eq:asymptsing_EB}) does indeed coincide with the numerically determined focal point of the spiral $\beta(E)$ for various arbitrary values $m$ in the allowed range (\ref{eq:asympt_mValid}) in dS and AdS. Even for values of $m$ outside the allowed range (\ref{eq:asympt_mValid}) (like $m=2$ and $m=-2$ of Figure \ref{fig:spirals}), the analytical expression gives a result very close to the numerical calculation.

\section{Homogeneous solution in dS}\label{sec:homo}

\begin{figure}[tb]
\begin{center}
	\includegraphics[scale=0.7]{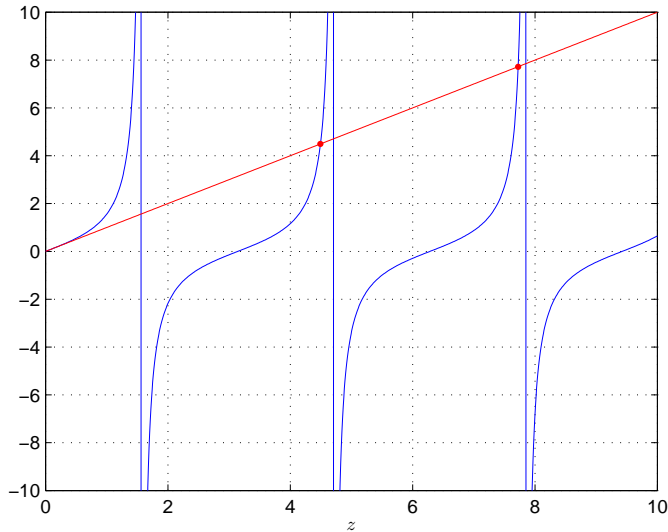} 
\end{center}
\caption{Graphical solution of the problem $\tan(z) = z$.}
\label{fig:tanz}
\end{figure}
\begin{figure}[tb]
\begin{center}
	\includegraphics[scale=0.7]{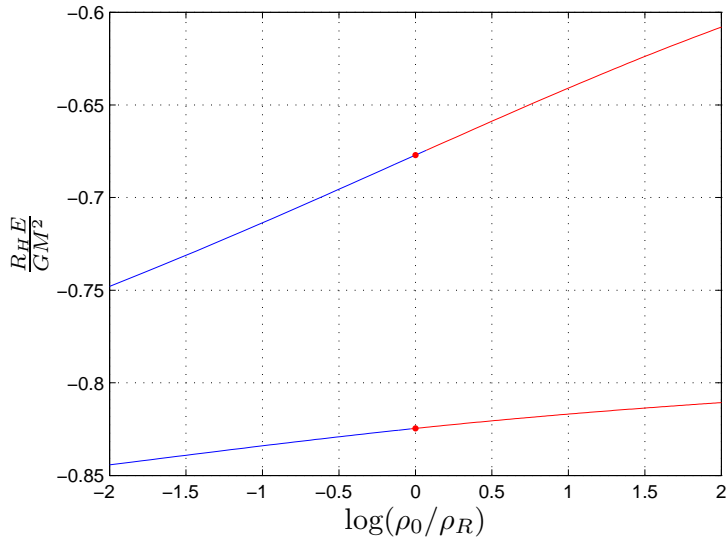} 
\end{center}
\caption{The energy versus the density contrast at the homogeneous radius $R = R_H$ for fixed $M$ and positive $\rho_\Lambda$. Unlike flat case, this plot \textit{cannot} be realized as keeping $E$ constant and varying $R$, because $m$ is held fixed. We see two distinct series of equilibria that are stable for $\rho_0 < \rho_R$ and unstable for $\rho_0 > \rho_R$. The change of stability happens at $\rho_0 = \rho_R$, which corresponds to specific homogeneous solutions, that indeed have a zero eigenvalue at these specific points ($ER/GM^2 = -0.677$ and $ER/GM^2 = -0.825$).}
\label{fig:RH}
\end{figure}

For a positive cosmological constant, the Emden-$\Lambda$ equation has solution with a uniform density (for $r<R$), we call homogeneous solution. For $\rho = const.$ the Poisson equation gives $\phi_N = \frac{2\pi G}{3}\rho r^2$, so that the full potential with the cosmological constant is
\begin{equation}\label{eq:phihom}
	\phi = \frac{2\pi G}{3}(\rho - 2\rho_\Lambda) r^2 + \phi(0)
\end{equation} 
The Poisson with $\Lambda$ equation (\ref{eq:poissonL}) gives $\phi = \phi(0) = const.$ for 
\[
	\rho = 2\rho_\Lambda
\] 
which of course is consistent with equation (\ref{eq:phihom}). This is the homogeneous solution with dimensionless temperature
\begin{equation}\label{eq:hombeta}
	B_H = \frac{1}{3} z^2
\end{equation}
and dimensionless energy
\begin{equation}\label{eq:homQ}
	Q_H = \frac{9}{2z^2} - \frac{9}{10}
\end{equation}
Since $\phi' =0$ the homogeneous solution has the same turning points in the two ensembles, because the eigenvalue equations (\ref{eq:eigen2}) and (\ref{eq:canonEig}) are the same in this case. Therefore, although the following analysis is done in terms of the microcanonical ensemble, the results hold for the canonical ensemble, as well. So that all turning points for the various solutions presented in this section are identical in the two ensembles. \\
\indent The radius $R_H$ of the homogeneous solution is independent of energy and temperature and is given for a fixed mass $M$ and cosmological constant $\rho_\Lambda$ by equation:
\begin{equation}\label{eq:Rhom}
	R_H = \left(\frac{3M}{8\pi\rho_\Lambda}\right)^{\frac{1}{3}}
\end{equation}
The homogeneous solution resembles the Einstein's static universe (\cite{Gibbons1,Gibbons2} has found the Einstein's static universe to be a local entropy maximum among other possible universes) in the Newtonian limit. However, to be more precise there are many homogeneous configurations (uniform density) with different temperatures, from which only one is literally static, that is the solution with $T=0$ ($\beta\rightarrow \infty$). The question is whether all, some or none of these solutions are stable. The static solution is surely thermodynamically unstable since it corresponds to only one microstate, therefore it has the minimum possible entropy, i.e. zero entropy. On the other hand the solution $\beta = 0$, which is allowed from (\ref{eq:hombeta}), is stable since it behaves as an ordinary gas. More formally, equation (\ref{eq:d2Sbeta}) can be written as 
\begin{equation}\label{eq:betad2S_1}
	\delta^{(2)}S = - \frac{\beta^2}{3M}\left( \int_0^R{dr \phi ' q} \right)^2
	+ \frac{G\beta}{2} \int_0^R dr \frac{q^2}{r^2} - \int_0^R dr\frac{{q'}^2}{8\pi \rho R^2}
\end{equation}
which gives $\delta^{(2)}S < 0$ for all perturbations $q$, for $\beta = 0$ ($T\rightarrow\infty$). In dimensionless variables equation (\ref{eq:betad2S_1}) reads
\begin{equation}\label{eq:betad2S_2}
	M\delta^{(2)}S = - \frac{1}{3}\left( \int_0^z{dx y' q} \right)^2
	+ \frac{1}{2}Bz \int_0^z dx \frac{1}{x^2}(q^2-{q'}^2 e^y)
\end{equation}
which gives $\delta^{(2)}S < 0$ for $B = 0$, as well. Therefore, there exists a point of change of stability somewhere between $B = 0$ and $B \rightarrow \infty$. From equation (\ref{eq:homQ}) we see that the energy has not an extremum. However, Poincar\'e's criterion does not exclude the possibility of having a change of stability at a point other than an energy extremum \cite{Katz1}. This is our case. The differential equation (\ref{eq:eigen3}) for the homogeneous solution $y=y'=0$ (and $m=\lambda = 1$) and for $\xi = 0$ becomes
\[
	F'' - \frac{2}{x}F' + F = 0
\]
which for $F(0) = 0$ has solution
\[
	F(x) = c(-x \cos(x) + \sin(x))
\]
We want the smallest $z$ that satisfies the second boundary condition $F(z) = 0$, that is the change of stability happens at this $z$ that is a solution of the equation
\[
	\tan(z) = z
\] 
The solution, call it $z_0$, can be found graphically (see Figure \ref{fig:tanz}) to obtain
\[
	z_0 \simeq 4.4934
\]
which corresponds to $B_0 \simeq 6.73$ from equation (\ref{eq:hombeta}). Therefore, the homogeneous solution is stable for temperature $T > T_0$ and unstable for $T < T_0$, with 
\begin{equation}\label{eq:homT0}
	T_0 \simeq \frac{GM}{6.73 R_H}
\end{equation}
\indent As we can see in Figure \ref{fig:m3D} there are infinite series of solutions for the homogeneous radius $R_H$ that corresponds to $m=1$ (the homogeneous solution corresponds to $m = \lambda = 1$). In Figure \ref{fig:RH} is drawn the dimensionless energy $Q = \frac{R_HE}{GM^2}$ versus the density contrast $\log(\rho_0/\rho_R)$ for two of these series. We see that there are solutions with $\rho_0 < \rho_R$ that continuously turn to solutions with $\rho_0 > \rho_R$. At the point $\rho_0 = \rho_R$, which corresponds to a homogeneous solution, there occurs a change of stability for the two solutions. That is because the corresponding energies $Q_0 = -0.6771$, $Q_1 = -0.8246$ correspond to the two first zero eigenvalues $z_0 = 4.4934$, $z_1 = 7.7251$ of the homogeneous solution as can easily be verified by equation (\ref{eq:homQ}). We numerically determined the unstable branch to be the one with $\rho_0 > \rho_R$. One is forced to conjecture that this pattern of different series, for $R=R_H$, is infinitely continued (as indicated by Figure \ref{fig:m3D}) for lower and lower energies. The change of stability at $\rho_0 = \rho_R$ should always correspond to $Q > - 0.9$ as indicated by equation (\ref{eq:homQ}). \\

\section{Microcanonical ensemble}\label{sec:microcanonical}
\begin{figure}[t!]
\begin{center}
	\subfigure[$R < R_H$]{ \label{fig:multa}\includegraphics[scale=0.4]{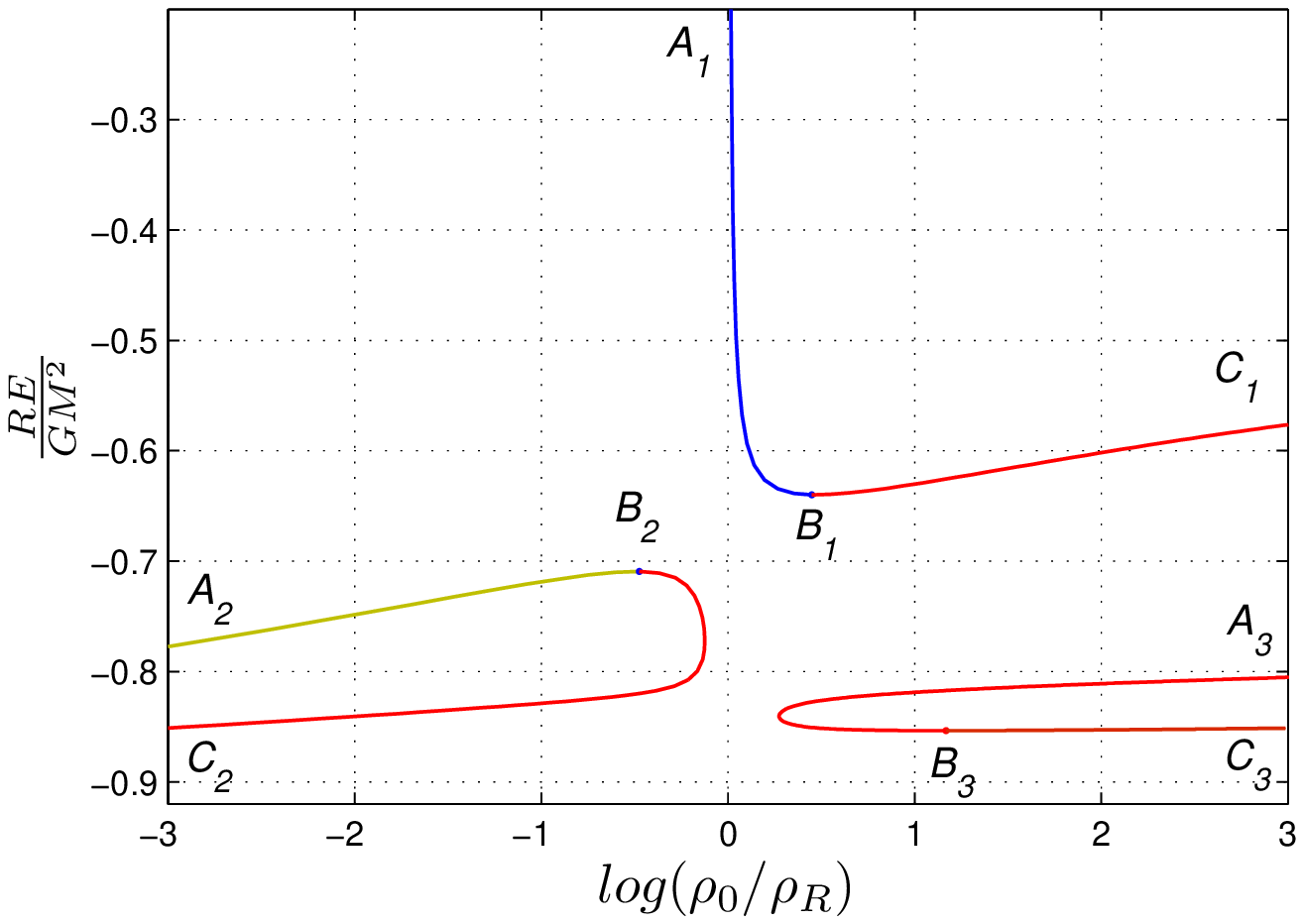} } 
	\subfigure[$R > R_H$]{ \label{fig:multb}\includegraphics[scale=0.4]{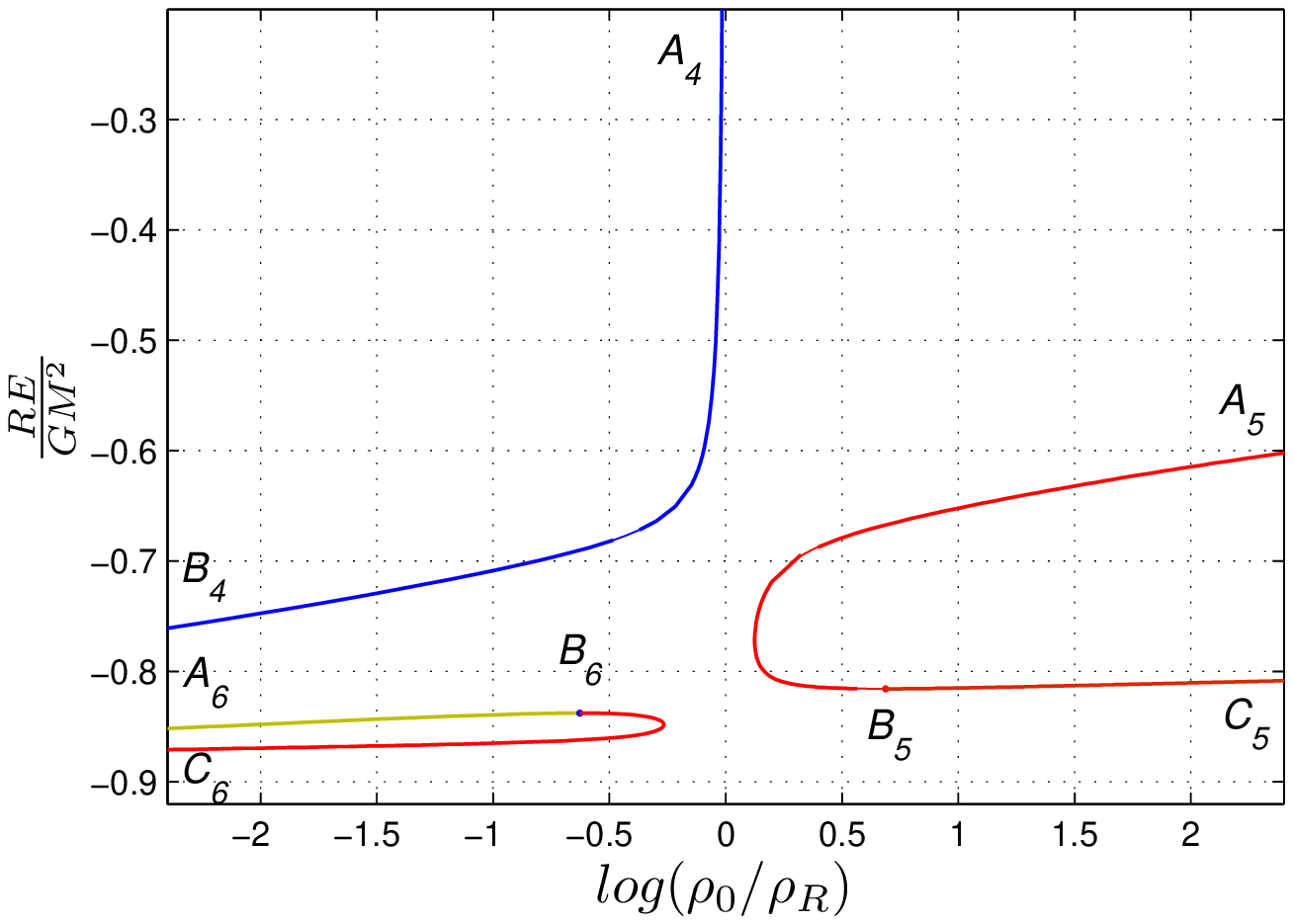} }
\caption{In dS, the dimensionless energy versus the density contrast for fixed $M$, $R$, $\rho_\Lambda$ in the microcanonical ensemble. Unlike flat case, this plot \textit{cannot} be interpreted with varying $R$ and fixed $E$, because at each point, $m$ is held constant. Only the upper series ($1$ and $4$) correspond to $\rho(r)$ monotonically changing. Distinct series corresponding to more negative energies have more extrema of $\rho(r)$. At points $B$, except $B_4$, an instability sets in. Curves $A_3B_3$ and $A_5B_5$ are already unstable solutions.
\label{fig:mult}}
\end{center}
\end{figure}

\indent In this section, we review our results of Ref. \cite{AGR}. Let investigate the solutions for $R<R_H$ and $R>R_H$ in the microcanonical ensemble. The energy versus the density contrast is plotted in Figure \ref{fig:mult}. The upper series labelled \textbf{1} ($R<R_H$) and \textbf{4} ($R>R_H$) correspond to monotonically changing density. For series \textbf{1} it is decreasing ($\rho_0 >\rho_R$), while for \textbf{4} increasing ($\rho_0 <\rho_R$). Series \textbf{1} suffers a change of stability at point $B_1$ (stable branch is $A_1B_1$), while series \textbf{4} is stable everywhere and does not suffer any change of stability. This is proved as follows: for this series the limit $E\rightarrow \infty$ does exist, which corresponds to $\beta = 0$. By equation (\ref{eq:betad2S_1}) for ${\beta} = 0$ we get $\delta^{(2)}S < 0$. In addition, the energy does not have an extremum (where a transition to instability could occur) and (to be sure) the whole series is numerically checked at each point for a zero eigenvalue. No one is found. Every such solution (series \textbf{4}) corresponds to configurations somewhat hollow at the center with matter concentrated mainly at the edge. The next series at more negative energies have one density extremum and at the next, one more is added and so on. At points $B$, except $B_4$ an instability sets in. For series $A_2B_2$ and $A_6B_6$ we have strong numerical evidence that are stable, while series \textbf{3} and \textbf{5} are found to be unstable. Series like $A_2B_2$ and $A_6B_6$ correspond to solutions diluted at the center with periodic condensations away from the center. One would normally expect this pattern in Figure \ref{fig:mult} to continue as one finds series with more and more negative energy. 

\subsection{Critical quantities}

\begin{figure}[tb]
\begin{center}
	\includegraphics[scale=0.5]{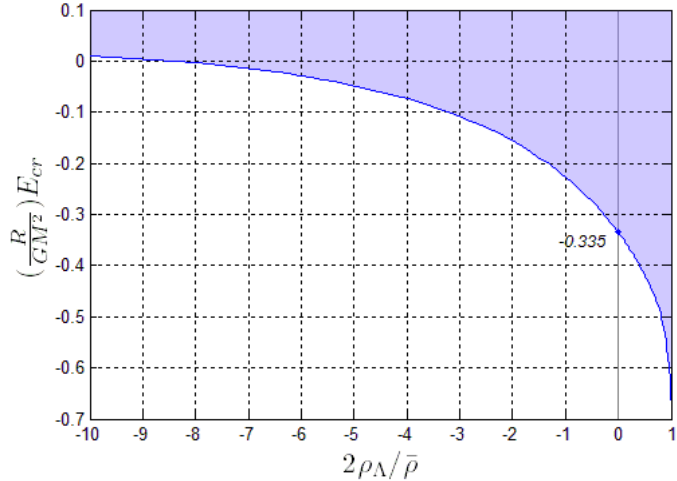} 
\end{center}
\caption{The critical energy versus $\rho_\Lambda$ for fixed $M$, $R$ in the microcanonical ensemble, where $\bar{\rho}$ is the mean density of matter. In the unshaded region there exist no equilibria. }
\label{fig:Ecr}
\end{figure}
\begin{figure}[tb]
\begin{center}
	\includegraphics[scale=0.7]{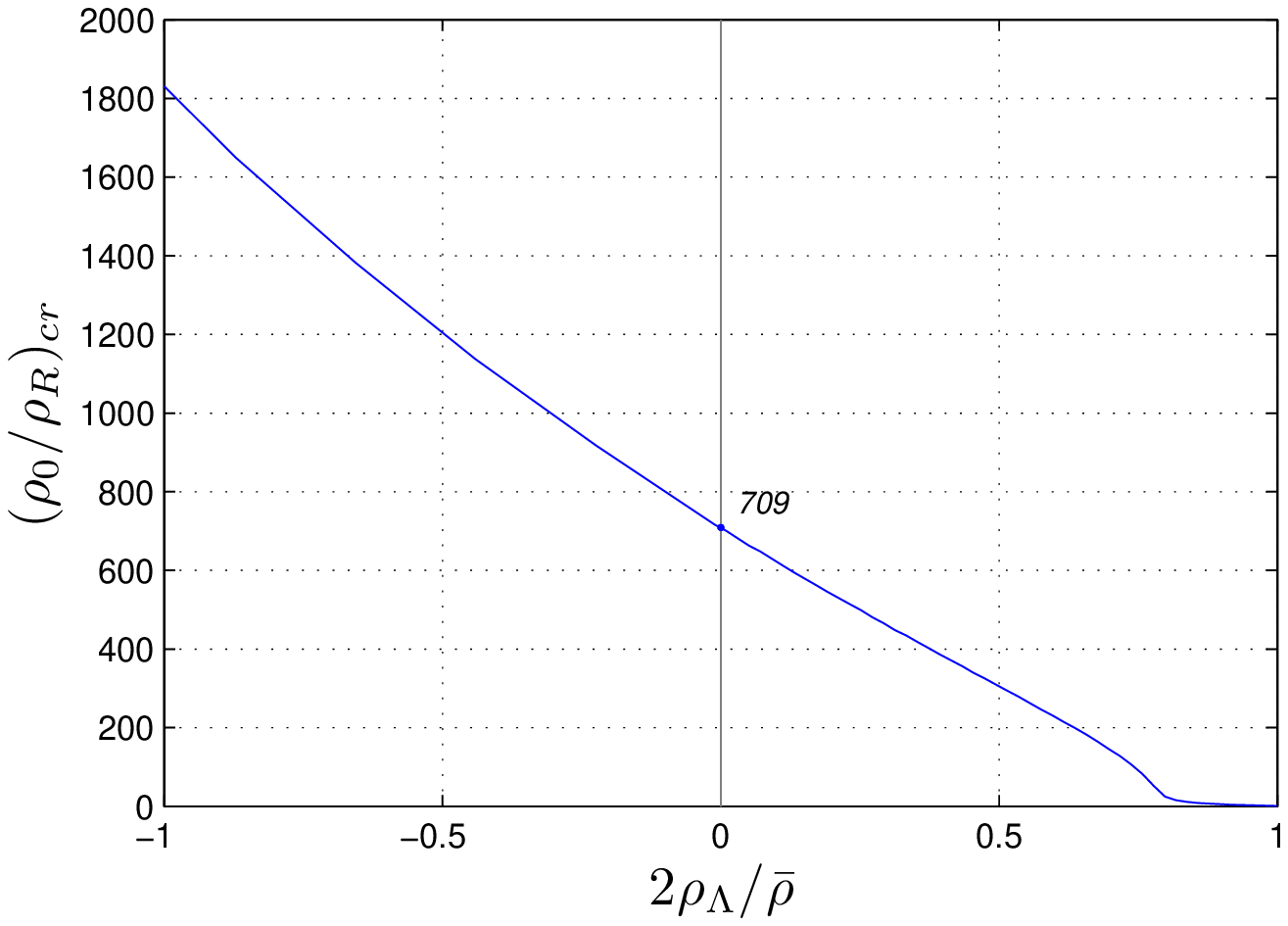} 
\end{center}
\caption{The critical density contrast versus $\rho_\Lambda$ for fixed $M$, $R$ in the microcanonical ensemble, where $\bar{\rho}$ is the mean density of matter. An instability sets in for $\rho_0/\rho_R > (\rho_0/\rho_R)_{cr}$ (except when $(\rho_0/\rho_R)_{cr} < 1$) where a clumping phase transition occurs. }
\label{fig:DCcr}
\end{figure}
\begin{figure}[tb]
\begin{center}
	\includegraphics[scale=0.6]{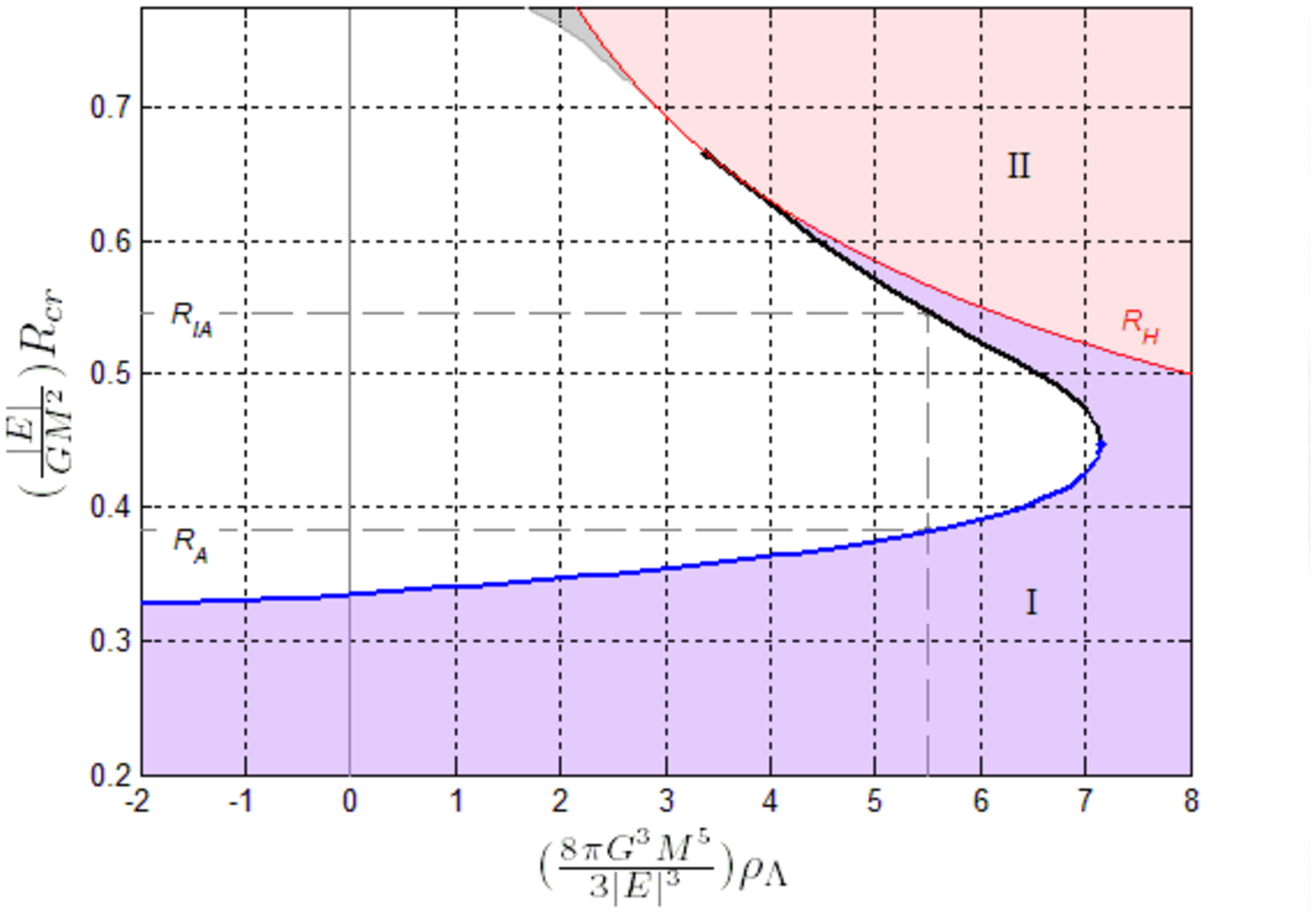} 
\end{center}
\caption{The critical radius versus $\rho_\Lambda$ for fixed $E$, $M$ in the microcanonical ensemble. There exist no equilibria in the unshaded region. With $R_H$ is denoted the radius of the homogeneous solution. See text for details.}
\label{fig:Rcr}
\end{figure}
A general result is that as the cosmological constant increases, isothermal spheres exist at even lower temperatures and energies. This can be seen in Figure \ref{fig:spirals}, where the classic spiral $\beta(E)$ is drawn for some positive and some negative cosmological constant. One could say that AdS destabilizes, while dS stabilizes the system.\\
\indent In the flat case, isothermal spheres exist only for 
\begin{equation}\label{eq:Ecrflat}
	E\cdot R > -0.335 GM^2
\end{equation}
This means that for a fixed radius, there exists a minimum critical energy $E_{cr} = -0.335 GM^2/R$ down to which, equilibria do exist. We want to determine how this critical energy changes with respect to $\rho_\Lambda$. The answer lies in Figure \ref{fig:Ecr}. The critical energy decreases for increasing cosmological constant. In AdS the critical energy becomes positive for $\rho_\Lambda \lesssim -4.2\bar{\rho}$, where $\bar{\rho}$ is the mean density.  \\
\indent Assuming the energy $E$ and radius $R$ are fixed at values that respect (\ref{eq:Ecrflat}) in flat case, does not guarantee that the equilibrium is stable. This depends on the density contrast $\rho_0/\rho_R$, that is the ratio of the central density versus the edge density. The situation is similar in the presence of $\rho_\Lambda$. For $\rho_\Lambda = 0$ it is $(\rho_0/\rho_R)_{cr} = 709$ \cite{Antonov} with the unstable branch being the one with $\rho_0/\rho_R > 709$. In Figure \ref{fig:DCcr} we see how this number changes w.r.t. the cosmological constant. The critical density contrast decreases for increasing cosmological constant. The instability occurs in AdS at more condensed configurations, while in dS at less condensed configurations. \\
\indent Assume that the energy is negative and fixed at some value $E$ in the flat case. Then, inequality (\ref{eq:Ecrflat}) shows that there is a maximum critical radius $R_A = (-0.335/E)GM^2$ that constrains the existence of an equilibrium. For $R>R_A$ there are no equilibria. In AdS this radius decreases as $\rho_\Lambda$ attains more negative values. In dS this radius increases as the cosmological constant increases and in addition there appears a second critical radius, we call $R_{IA}$, that constrains the existence of equilibria from below. At $R_{A}$ and $R_{IA}$ a collapse phase transition takes place. That is, no equilibria exist for $R_A < R < R_{IA}$ as can be seen in Figure \ref{fig:Rcr} and the system lies in a collapsed phase. This is a typical reentrant behavior, that is common to statistical systems \cite{reentrant1,reentrant2,reentrant3,reentrant4}, whenever competing interactions are present. Beyond the marginal value $\rho_\Lambda^{dS} \simeq 7.14 (3|E|^3/8\pi G^3 M^5)$ there can always be found equilibrium states in dS case. \\
\indent In region I of Figure \ref{fig:Rcr} there exist series \textbf{1} of equilibria of Figure \ref{fig:multa} and in region II all equilibria of Figure \ref{fig:multb}. In the small upper gray shaded region there are the rest series of Figure \ref{fig:multa}. These type of equilibria exist only for values of the cosmological constant greater than a minimum value $\rho_\Lambda^{min}$. This is the smallest value for which the cosmological force can keep marginally the whole matter at the edge. It can easily be calculated by equating the forces at the edge, assuming all matter is concentrated at $R$:
\begin{equation}\label{eq:lambdamin}
	\frac{GM^2}{2R} = G M \frac{8}{3}\pi \rho_\Lambda^{min} R^2 \Rightarrow
	\rho_\Lambda^{min} = \bar{\rho}/4
\end{equation}
where $\bar{\rho}$ is the mean density.

\subsection{Comparison with Schwartzschild-dS space}\label{sec:SdS}

\begin{figure}[tb]
\begin{center}
	\includegraphics[scale=0.78]{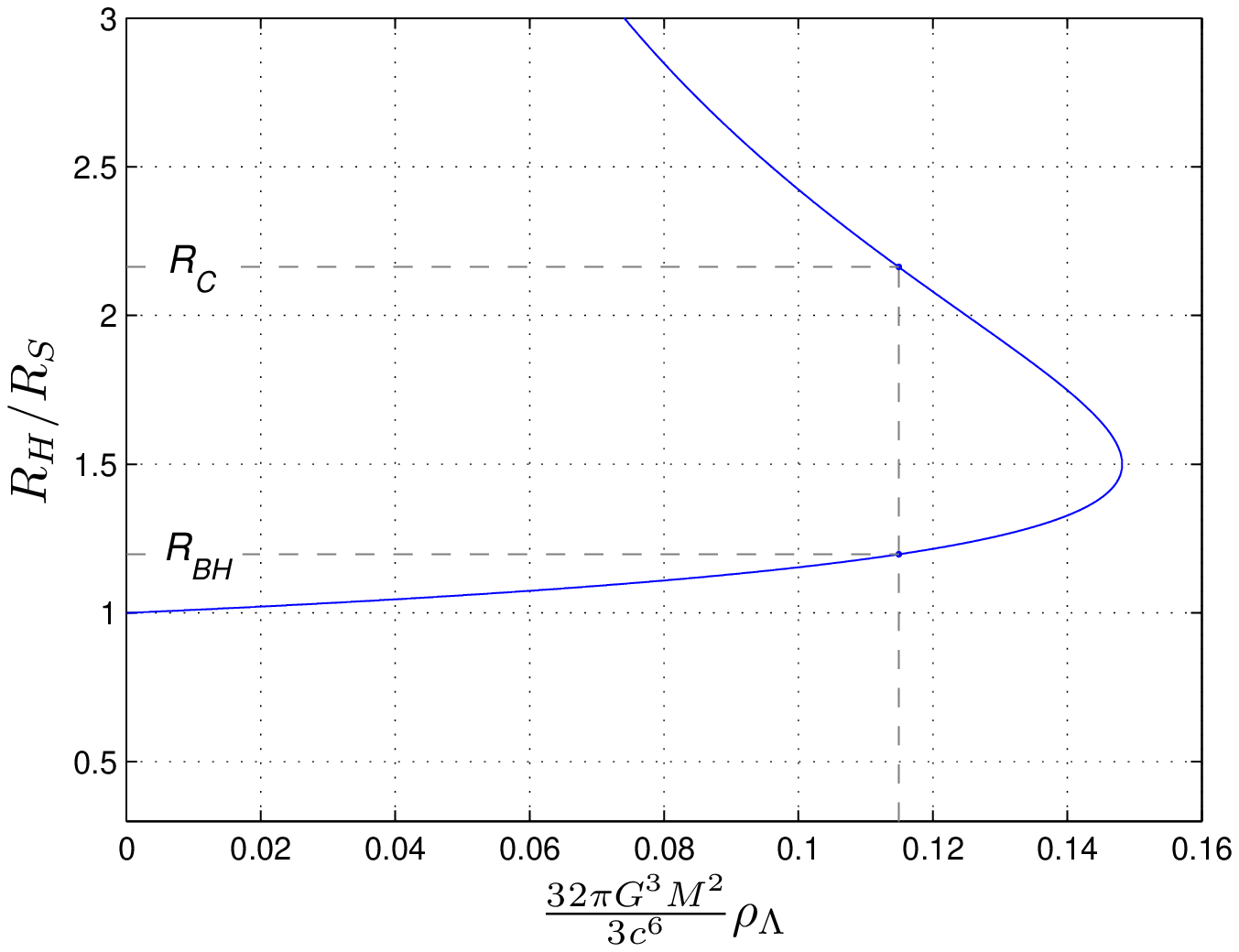}  
\end{center}
\caption{The horizons of the Schwartzschild-dS space versus the cosmological constant for a fixed mass. The horizon radius $R_H$ is measured in units of the Schwartzschild radius $R_S = \frac{2GM}{c^2}$. For a given cosmological constant there are two horizons: the black hole horizon $R_{BH}$ and the cosmological horizon $R_C$. The similarity with Figure \ref{fig:Rcr} is striking!}
\label{fig:SdS}
\end{figure}
The two critical radii in Figure \ref{fig:Rcr} resemble the two horizons of Schwartzschild-dS space, where the role of the cosmological horizon is played by $R_{IA}$. The Schwartzschild-dS metric can be written as:
\begin{equation}\label{eq:SchwdS}
	ds^2 = - \left( 1-\frac{2GM}{c^2r}-\frac{8\pi G}{3c^2}\rho_\Lambda r^2 \right)dt^2
			+ \left( 1-\frac{2GM}{c^2r}-\frac{8\pi G}{3c^2}\rho_\Lambda r^2 \right)^{-1}dr^2 + r^2d\Omega_2
\end{equation}
where $\rho_\Lambda$ is the `mass' density of the cosmological constant (the energy density is $\kappa = \rho_\Lambda c^2 = \frac{\Lambda c^4}{8\pi G}$). This metric has two horizons for $\rho_\Lambda > 0$; the black hole horizon $R_{BH}$ and the cosmological horizon $R_C$. Both are defined as the real roots $R_H$ of the third order polynomial equation
\begin{equation}\label{eq:hor}
	1-\frac{2GM}{c^2R_H}-\frac{8\pi G}{3c^2}\rho_\Lambda R_H^2 = 0
\end{equation}
which, for various values of the cosmological constant $\rho_\Lambda$, are plotted in Figure \ref{fig:SdS}. The resemblance with Figure \ref{fig:Rcr} is too much striking to be considered accidental! It seems as though the reentrant phenomenon of Figure \ref{fig:Rcr} is the closest Newtonian analogue of the horizons (Figure \ref{fig:SdS}) in Schwartzschild-dS space. \\
\indent However, there is a big deference between the Schwartzschild-dS space and the Newtonian reentrant phenomenon. It is the opposite sense of the inequality for the instability, i.e. the stable region in the Newtonian case corresponds to the unstable region ($R < R_{BH}$ and $R > R_C$) of the Schwartzschild-dS space. \\

\section{Canonical ensemble}\label{sec:canonical}
\begin{figure}[here!]
\begin{center}
	\subfigure[$R < R_H$]{ \label{fig:Tmulta}\includegraphics[scale=0.4]{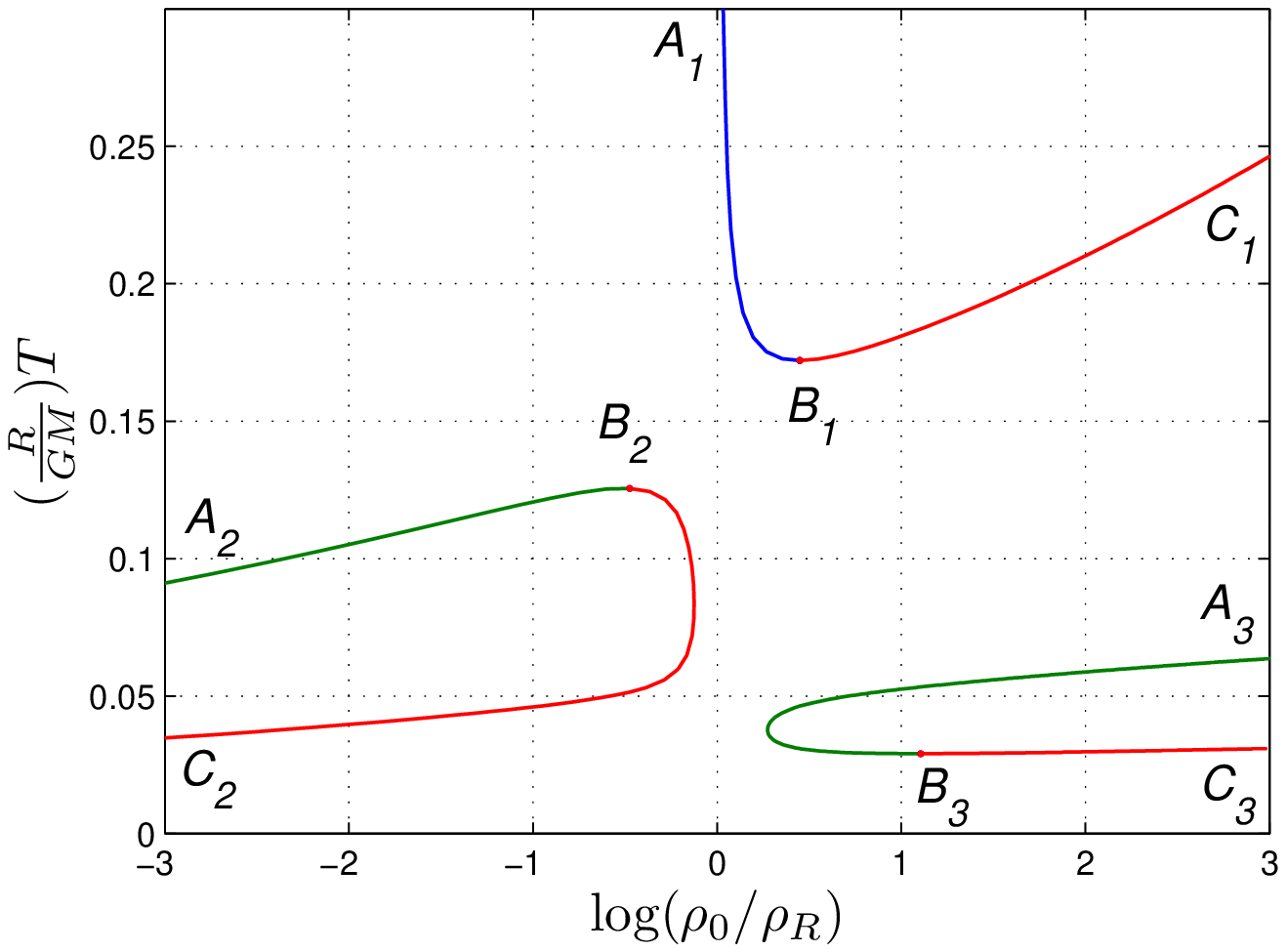} } 
	\subfigure[$R > R_H$]{ \label{fig:Tmultb}\includegraphics[scale=0.4]{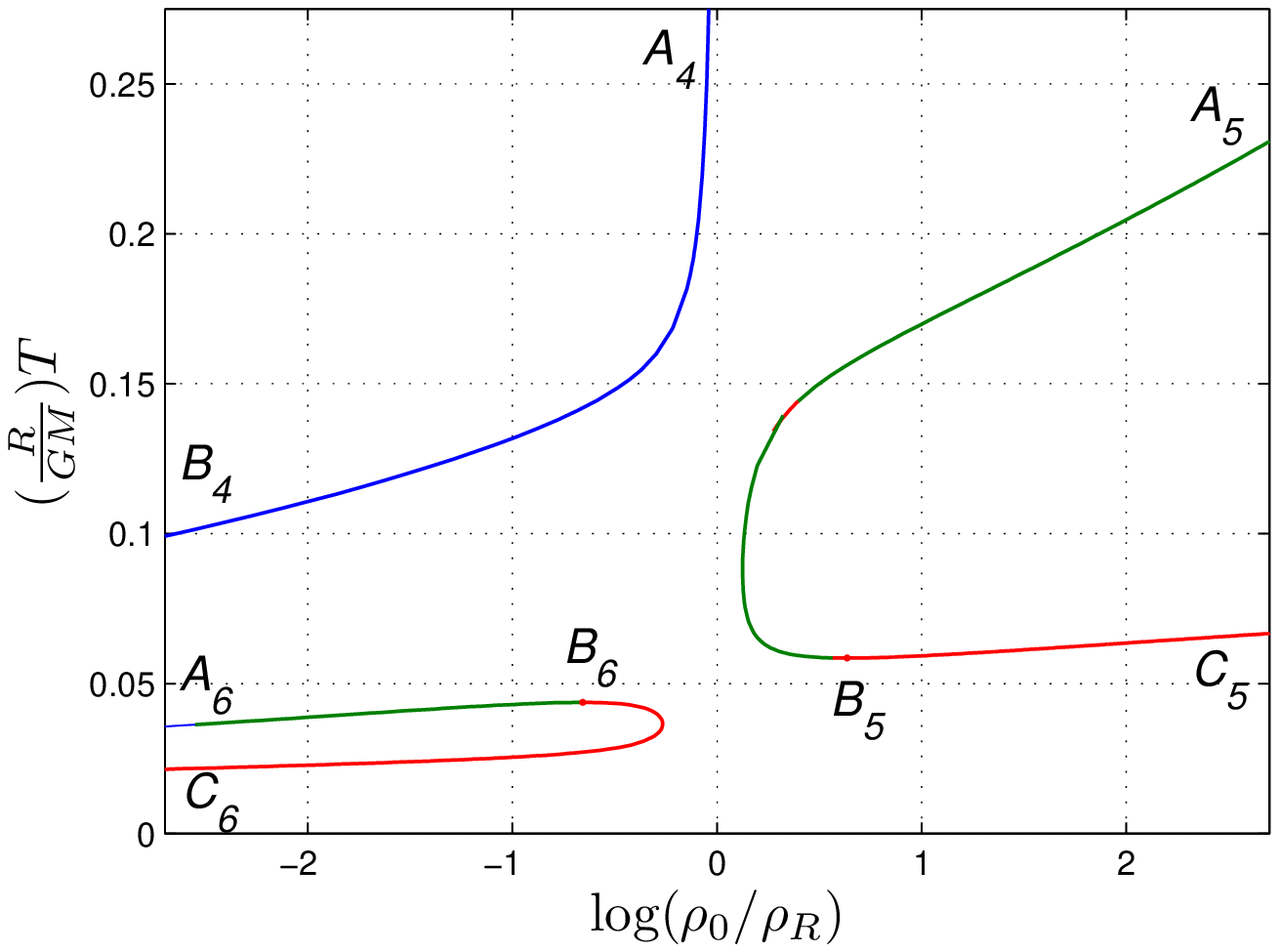} }
\caption{In dS, the dimensionless temperature versus the density contrast for fixed $M$, $R$, $\rho_\Lambda$ in the canonical ensemble. Only the upper series ($1$ and $4$) correspond to $\rho(r)$ monotonically changing. At points $B$, except $B_4$, an instability sets in. See text for details.
\label{fig:Tmult_C}}
\end{center}
\end{figure}

\begin{figure}[tb]
\begin{center}
	\includegraphics[scale=0.5]{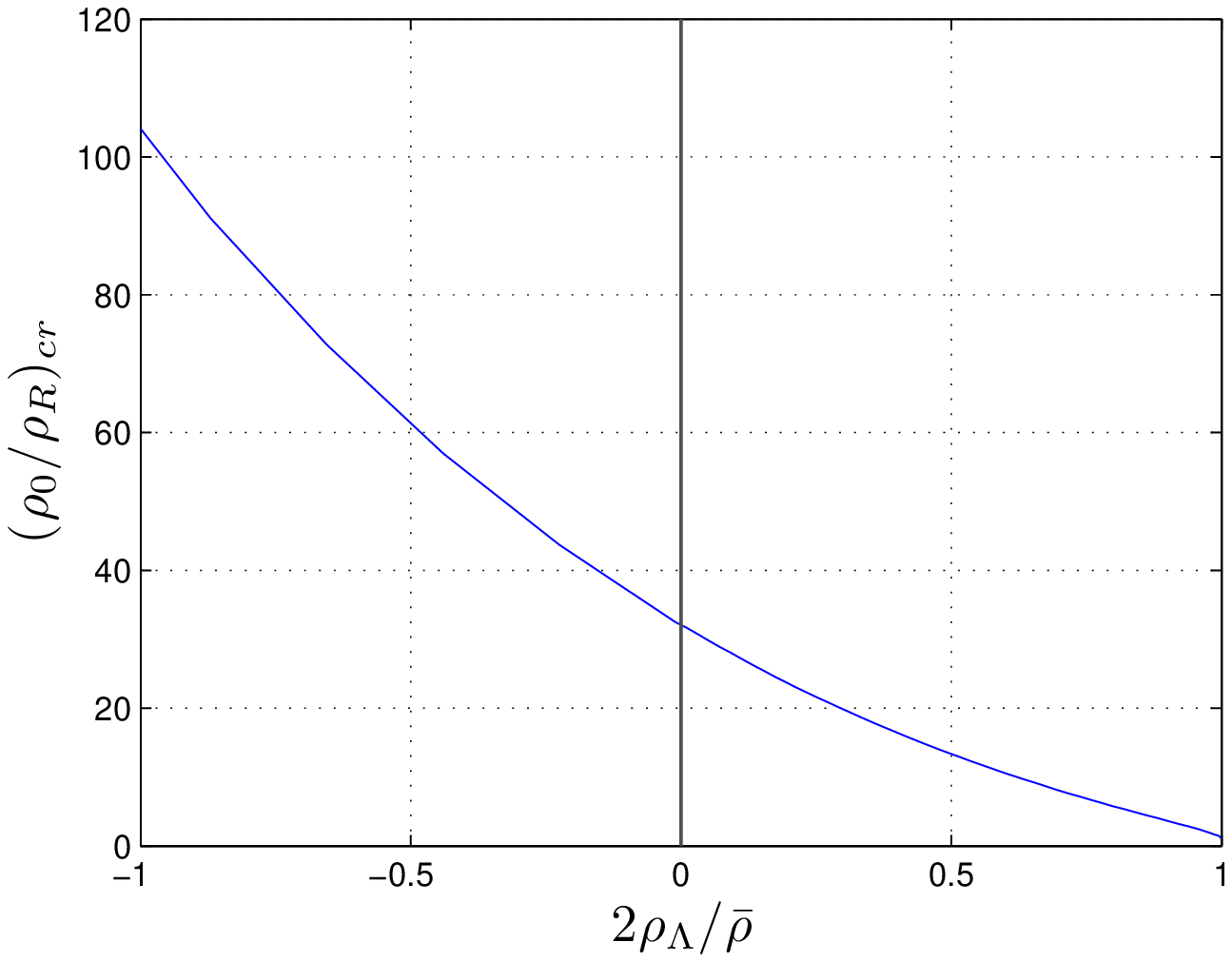} 
\end{center}
\caption{The critical density contrast versus $\rho_\Lambda$ for fixed $M$, $R$ in the canonical ensemble, where $\bar{\rho}$ is the mean density of matter. An instability sets in for $\rho_0/\rho_R > (\rho_0/\rho_R)_{cr}$ (except when $(\rho_0/\rho_R)_{cr} < 1$), where a clumping phase transition occurs. }
\label{fig:DCcr_C}
\end{figure}

\begin{figure}[tb]
\begin{center}
	\includegraphics[scale=0.6]{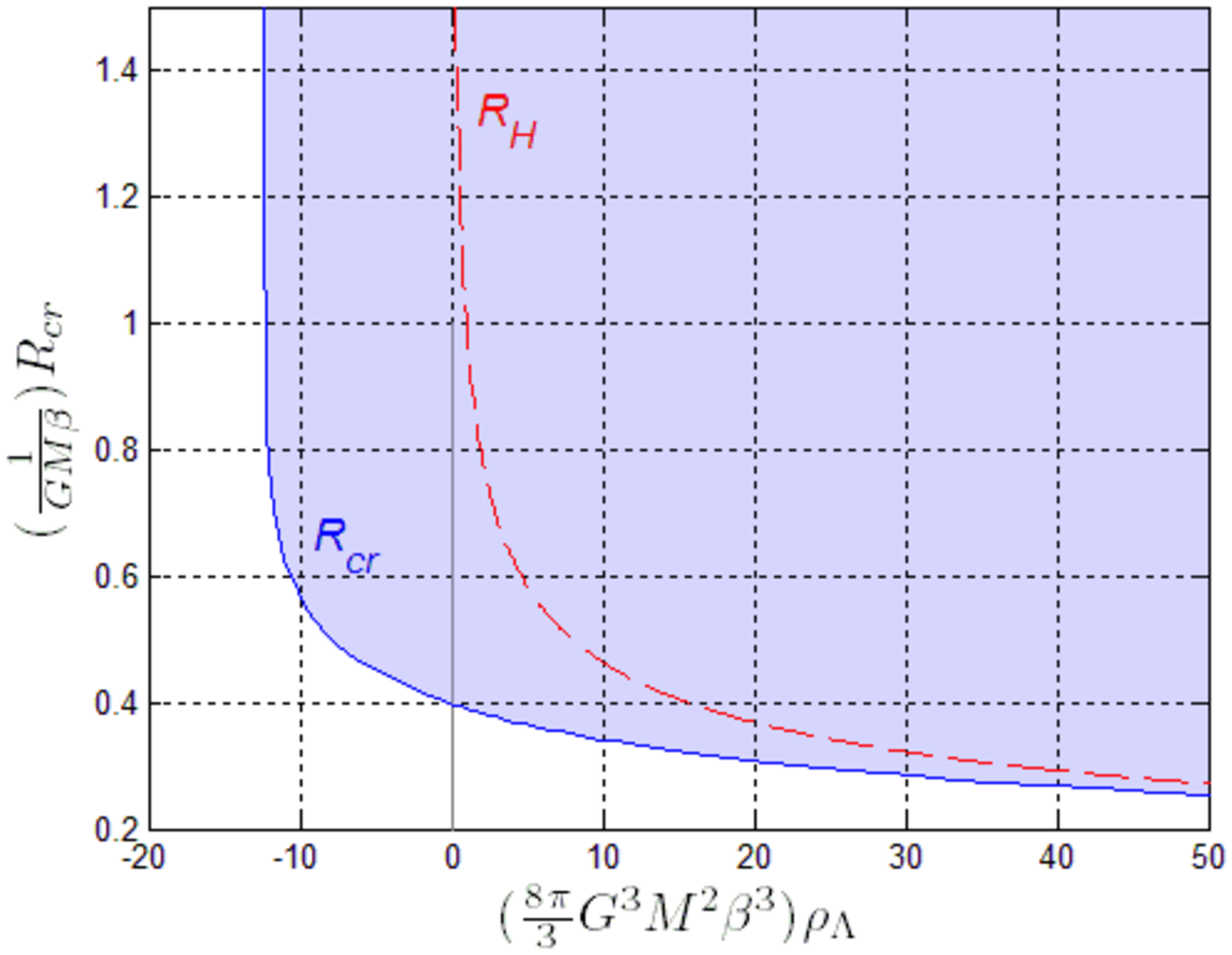} 
\end{center}
\caption{The critical radius versus $\rho_\Lambda$ for fixed $\beta$, $M$ in the canonical ensemble. There exist no equilibria in the unshaded region. With $R_H$ is denoted the radius of the homogeneous solution. See text for details.}
\label{fig:Rcr_C}
\end{figure}

\begin{figure}[tb]
\begin{center}
	\includegraphics[scale=0.6]{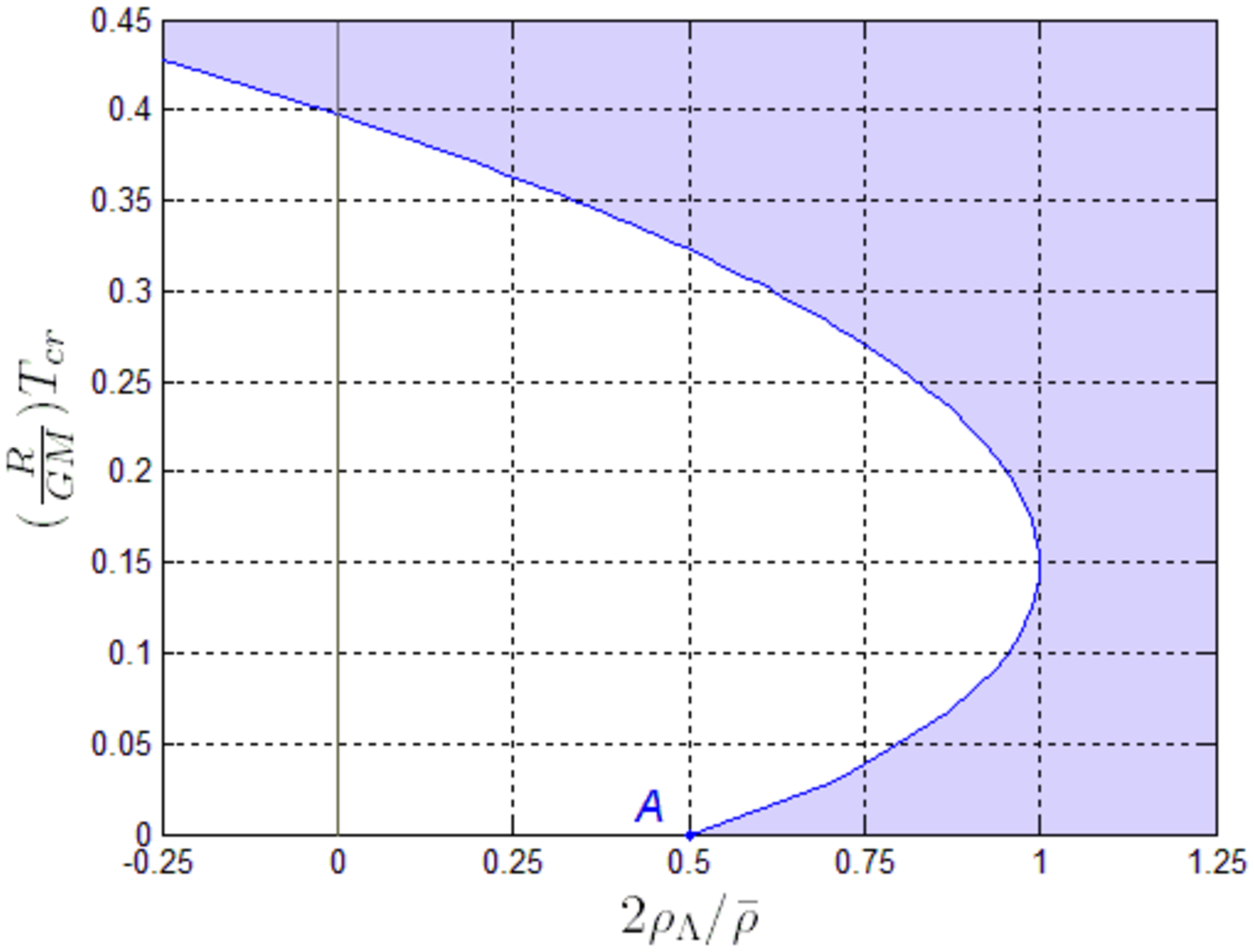} 
\end{center}
\caption{The critical temperature versus $\rho_\Lambda$ for fixed $M$, $R$ in the canonical ensemble, where $\bar{\rho}$ is the mean density of matter. In the unshaded region there exist no equilibria. This behavior indicates a reentrant collapse phase transition. }
\label{fig:Tcr_C}
\end{figure}

\indent In the canonical ensemble \cite{AGR2} the instability sets in at different points than in the microcanonical ensemble. The series of equilibria expressed as function of temperature with respect to the density contrast can be seen in Figure \ref{fig:Tmult_C} in case of dS. In AdS, just like flat case, there exist only one series of equilibria for a fixed cosmological constant, that is similar in form to series \textbf{1} of Figure \ref{fig:Tmulta}. We note that Figure \ref{fig:Tmult_C} of temperature is very similar to 
Figure \ref{fig:mult} of the energy. At points $B$, except $B_4$, of Figure \ref{fig:Tmult_C}, an instability sets in. Series $A_1B_1$ and $A_4B_4$ are stable. A safe conclusion on the stability of the green series $A_2B_2$, $A_3B_3$, $A_5B_5$ and $A_6B_6$ could not be reached, although the positive specific heat indicates stability, since we work in the canonical ensemble. The rest series are unstable. \\
\indent In Figure \ref{fig:DCcr_C} we can see that the critical density contrast for series \textbf{1} of Figure \ref{fig:Tmulta}, is decreasing for increasing cosmological constant, just like in the microcanonical ensemble. However, the instability sets in earlier in the canonical ensemble. It is well known in the flat case, and this fact is true with a cosmological constant, too, that the instability in the canonical ensemble sets in when the specific heat becomes negative. This negative specific heat region is stable in the microcanonical ensemble, while the instability sets in when the negative specific heat becomes positive again, in this ensemble (see \cite{Bell-Wood}). \\
\indent The canonical ensemble is completely different then the microcanonical ensemble as far as the stability of the system is concerned. We alway consider the mass to be fixed in both ensembles. As can be seen in Figure \ref{fig:Rcr_C}, the critical radius is decreasing with increasing cosmological constant and the region of the instability is now for radii smaller (and not bigger) then the critical radius $R < R_{cr}$. The critical radius is decreasing for increasing $\rho_\Lambda$, because for an increase in $\rho_\Lambda$ the temperature is decreased, so that one should compress the system to balance out this destabilizing temperature decrease. The region of the instability changes, because in a compression although the pressure gradient is increased, it is weakened compared to the microcanonical ensemble, due to the heat transferred to the heat bath. In AdS case there is a marginal value of the cosmological constant 
$\rho_\Lambda^{AdS} \simeq -12.32 (3/8\pi G^3M^2\beta ^3)$ beyond which no equilibrium states are possible. \\
\indent The critical temperature is decreasing with increasing cosmological constant, as can be seen in Figure \ref{fig:Tcr_C}. Increase of the cosmological constant acts as a stabilizer on the system due to the increase of the repelling force (or the decrease of the attracting force in case of AdS), enabling the system to be stable at lower temperatures. As the cosmological constant increases, it reaches a value for which the system can marginally be in \textit{static} dynamical equilibrium. At this state all matter is still, i.e. $T = 0$, and is concentrated at the edge. This is point $A$ in Figure \ref{fig:Tcr_C}. We have calculated this point earlier in equation (\ref{eq:lambdamin}) and found $\rho_\Lambda^{min} = \bar{\rho}/4$. For greater values the outward pointing cosmological force is increasing, enabling the mass to approach towards the center and to greater temperatures. The system undergoes a reentrant phase transition in the canonical ensemble. For a fixed cosmological constant at this region (after point $A$), there are metastable states for low temperatures up to some maximum critical value $T_1$, where a collapse phase transition occurs. For greater temperature, there exist no metastable states and the system suffers isothermal collapse. This happens up to some second critical temperature $T_2$. For even greater temperatures, the equilibria are restored. 

\section{Conclusions}

In the presence of the cosmological constant we find that, in the microcanonical ensemble, the critical radius, namely the Antonov radius, above which the system becomes unstable is increasing with increasing (negative or positive) cosmological constant, while in the canonical ensemble it is decreasing. The critical energy and temperature (less than which the system becomes unstable), as well as the critical density contrast $(\rho_0/\rho_R)_{cr}$, are decreasing in the two ensembles. \\
\indent In dS case a new phenomenon is discovered, namely a reentrant phase transition; in microcanonical ensemble there emerges a second critical radius above which metastable states are restored, while in the canonical there appears a second critical low temperature, lower than which equilibria are restored. We stress out the similarity of the behavior, we have discovered, of the two critical radii in dS case in the microcanonical ensemble with the two horizons of relativistic Sschwartzscild-dS space. It seems that our `dS case' is a Newtonian analogue of Schwartzschild-dS system. \\
\indent Another interesting feature of dS is the turning point of stability for the homogeneous solution, which resembles the Einstein's static universe, and the infinite numbers of non-uniform solutions at the homogeneous radius which suffer a transition from stability to instability when passing from solutions with density contrast $\rho_0/\rho_R < 1$ to the ones with $\rho_0/\rho_R > 1$. In addition, there exist multiple series of equilibria for a given positive cosmological constant and fixed radius and mass. \\
\indent We stress out, that in this work the non-equivalence of ensembles in gravitating systems is confirmed in the most dramatic way. For a fixed mass, in the microcanonical ensemble the instability occurs for radii larger than a critical value, while in the canonical the instability occurs for radii smaller than a critical value. That is because in the canonical ensemble, the pressure gradient that balances gravitation, is not drastically increased during a compression of the system, due to the heat transfer to the heat bath. In contrast, in the microcanonical ensemble where there is no energy loss, the pressure gradient is increased during a compression, drastically enough to balance gravity. \\
\indent Regarding applications to the physical world (positive cosmological constant), we state two interesting issues raised by our work. In the microcanonical ensemble, our analysis could have implications to the evolution of galaxy clusters, since many of them present a core in their centre and others a supermassive black hole. A quick calculation shows that, for the observed value of $\rho_\Lambda$, the relevant to our stability analysis, dimensionless quantities $2\rho_\Lambda/\bar{\rho}$ and $8\pi G^3 M^5\rho_\Lambda/3|E|^3$ are of order unity for some typical values of regular galaxy clusters \cite{AGR}. This implies that the cosmological constant could influence the onset of the instability of galaxy clusters. In the canonical ensemble, the cosmological constant could have an effect on the fractal structure of the Universe, since it is connected with the secondary instabilities \cite{Chavanis1}. Our analysis on the asymptotic behavior of the equilibria (section \ref{sec:series}), can be used to calculate the secondary instabilities. \\
\indent Let us close, noting that, even though there might be phenomenological connection with the physical world, our work is mainly focused and wishes to contribute on the theoretical understanding of the impact of an arbitrary cosmological constant term to the stability of gravitational systems.


\appendix

\section{Antonov's proof that global entropy maxima do not exist}\label{app:A}
\begin{figure}[here!]
\begin{center}
\psset{unit=0.8cm}
\begin{pspicture}(-2,-1)(2,7)
	\psset{Alpha=58,Beta=20}
 \pstThreeDSphere[SegmentColor={[cmyk]{0,0,0,0}}](0,0,4){2}
 \pstThreeDCoor[xMax=4,yMax=4,zMax=7]
 \pstThreeDDot[drawCoor=false,linecolor = red](1.5,1.7,4)
 \pstThreeDLine[linestyle=dashed](0,0,0)(1.5,1.7,0)
 \pstThreeDLine[linestyle=dashed](1.5,1.7,0)(1.5,1.7,4)
 \pstThreeDCircle[beginAngle=0,endAngle=50,arrows=->](0,0,0)(0.8,0,0)(0.5,0.5,0)
 \pstThreeDPut[pOrigin=lb](0.8,0.9,0){$\tau$}
 \pstThreeDPut[pOrigin=ct](1.3,0.6,0){$\phi$}
 \pstThreeDPut[pOrigin=l](1.5,1.8,2){$z$}
 \pstPlanePut[plane=yz](0,3,0){\pstThreeDLine[arrows=|-|,linecolor = gray](0,0,0)(0,0,4)}
 \pstPlanePut[plane=yz](0,3.1,2){$r_{12}$}
 \pstPlanePut[plane=yz](0,3.4,0){\pstThreeDLine[arrows=|-|,linecolor = gray](0,0,4)(0,0,6)}
 \pstPlanePut[plane=yz](0,3.5,5){$r_{2}$}
 \pstThreeDLine[linestyle=dashed,linecolor = gray](0,0,4)(0,3.4,4)
 \pstThreeDLine[linestyle=dashed,linecolor = gray](0,0,6)(0,3.4,6)
 \pstThreeDDot[](0,0,4)
 \pstThreeDLine[arrows=->](0,0,0)(1.5,1.7,4)
 \pstThreeDPut[pOrigin=l](0,0.3,1){$\vec{r}$}
\end{pspicture}
\end{center}
\caption{The cylindrical coordinates for the second sphere. The first sphere, which is not drawn, has its center at the origin.} 
\label{fig:coord}
\end{figure}
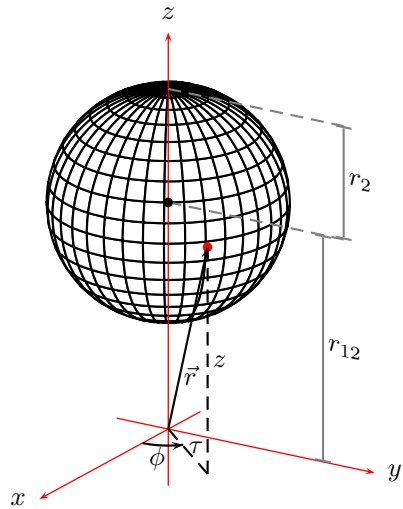
Let $0<\alpha < 1$ and $\alpha M$ be the mass uniformly distributed inside a sphere with radius $r_1$ and $(1-\alpha)M$ the mass uniformly distributed inside a second sphere with radius $r_2$. The distance of the spheres' centers is $r_{12}$. The distribution functions are constants
\[
	f_1 = \frac{\alpha M}{\frac{4}{3}\pi r_1^3\frac{4}{3}\pi \upsilon_1^3} = \frac{\alpha M}{\frac{16}{9}\pi^2 r_1^3\upsilon_1^3}
	\; , \;
	f_2 = \frac{(1-\alpha) M}{\frac{16}{9}\pi^2 r_2^3\upsilon_2^3}
\]
where $\upsilon_1$, $\upsilon_2$ the corresponding to the two spheres velocity bounds and the velocities $\upsilon$ are equally probable by construction. \\
Let us calculate the entropy of the system:
\begin{eqnarray}\label{eq:appA1}
	S = &-&k\alpha M \log f_1 - k(1-\alpha)\log f_2 = -kM\lbrace\alpha \log\alpha + (1-\alpha)\log(1-\alpha) \rbrace \nonumber \\
	 &+& 3kM\lbrace(1-\alpha)\log \upsilon_2 r_2 + \alpha\log\upsilon_1 r_1\rbrace
	-kM\log\frac{9M}{19\pi^2} 
\end{eqnarray}
The Newtonian dynamical energy $U_1$ of the first sphere is
\[
	U_1 = -\int_0^{r_1} G\frac{\rho\frac{4}{3}\phi r^3}{r}dm = -GM^2\alpha^2\frac{3}{r_1^6}\int_0^{r_1} r^4 dr = 
	-\frac{3}{5}GM^2\frac{\alpha^2}{r_1}
\]
and similarly for the second sphere
\[
	U_2 = -\frac{3}{5}GM^2\frac{(1-\alpha)^2}{r_2}
\]
Let our coordinate system be centered at the center of the first sphere and the center of the second sphere be at $z=r_{12}$. 
The cosmological dynamical energy for the first sphere is
\[
	U_{\Lambda 1} = \int_0^{r_1}\rho\phi_\Lambda d^3\vec{r} = -\frac{4\pi G}{3}\rho_\Lambda \frac{\alpha M}{\frac{4}{3}\pi r_1^3}\int_0^{r_1}4\pi r^4 dr = -\frac{4}{5}\pi G\rho_\Lambda M \alpha r_1^2
\]
For the second sphere we use the cylindrical coordinates $(\tau,\phi,z)$ (see Figure \ref{fig:coord}). Its cosmological dynamical energy is
\begin{eqnarray}
	U_{\Lambda 2} &=& \int_V\rho\phi_\Lambda d^3\vec{r} = -\frac{4\pi G}{3}\rho_\Lambda \frac{(1-\alpha)M}{\frac{4}{3}\pi r_2^3}
	\cdot 2\int_0^{2\pi} d\phi\int_{r_{12}-r_2}^{r_{12}}\left\lbrace\int_0^{r_2}(z^2 + \tau^2)\tau d\tau\right\rbrace dz 
	\nonumber \\
	&=& -4\pi G\rho_\Lambda M (1-\alpha)\left\lbrace \frac{1}{6}\frac{1}{r_2}(r_{12}^3 - (r_{12}-r_2)^3)+\frac{1}{4}r_2^2\right\rbrace \nonumber
\end{eqnarray}
The Newtonian dynamical energy between the two spheres is
\[
	U_{12} = -G\frac{M^2\alpha(1-\alpha)}{r_{12}}
\]
The kinetic energy of the first sphere is
\[
	K_1 = \frac{1}{2}\int f_1\upsilon^2 d^3\vec{\upsilon} d^3\vec{r} = \frac{1}{2}\int_0^{\upsilon_1}\int_0^{r_1} 
	\frac{\alpha M \upsilon^2}{\frac{16}{9}\pi^2 r_1^3\upsilon_1^3}4\pi \upsilon^2d\upsilon\, 4\pi r^2 dr = 
	\frac{3}{10}\alpha M \upsilon_1^2
\]
and similarly for the second sphere
\[
	K_2 = \frac{3}{10} (1-\alpha)M\upsilon_2^2
\]
Summing it all up, the total energy is
\begin{eqnarray}\label{eq:appA2}
	E = &-& \frac{3}{5}GM^2\left\lbrace \frac{\alpha^2}{r_1} + \frac{(1-\alpha)^2}{r_2} + \frac{5}{3}\frac{\alpha(1-\alpha)}{r_{12}}\right\rbrace \nonumber \\
	&-& \frac{4\pi G}{5}M\rho_\Lambda\left\lbrace \alpha r_1^2 + (1-\alpha) \left[ \frac{5}{6}\frac{1}{r_2}(r_{12}^3 - (r_{12}-r_2)^3)+\frac{5}{4}r_2^2 \right] \right\rbrace\nonumber \\
	&+& \frac{3}{10}M\left\lbrace \alpha \upsilon_1^2 + (1-\alpha) \upsilon_2^2\right\rbrace 
\end{eqnarray}
We keep constant the quantities $r_1$, $\upsilon_2$ and $L = -(1-\alpha)\log\upsilon_2 r_2$. Equation (\ref{eq:appA2}) gives
\begin{eqnarray}\label{eq:appA3}
	\upsilon_1^2 = && \frac{1}{\alpha}\left\lbrace\frac{10}{3M}\left( E + \frac{GM^2\alpha(1-\alpha)}{r_{12}}\right) + 2GM\frac{\alpha^2}{r_1} - (1-\alpha)\upsilon_2^2 + \frac{2GL^2 M}{r_2 (\log\upsilon_2 r_2)^2} \right. \nonumber\\
	 &+& \left. \frac{8\pi G}{3}\rho_\Lambda \left(\alpha r_1^2 + (1-\alpha) \left[ \frac{5}{6}\frac{1}{r_2}(r_{12}^3 - (r_{12}-r_2)^3)+\frac{5}{4}r_2^2 \right]\right)\right\rbrace
\end{eqnarray} 
For $r_2\rightarrow 0$ equation (\ref{eq:appA2}) gives $\alpha\rightarrow 1$ in order to keep the energy finite. 
For $r_2\rightarrow 0$, $\alpha\rightarrow 1$ equation (\ref{eq:appA3}) gives $\upsilon_1 \rightarrow \infty$.
For $r_2\rightarrow 0$, $\alpha\rightarrow 1$ and $\upsilon_1 \rightarrow \infty$ equation (\ref{eq:appA1}) gives
$S\rightarrow \infty$. Therefore there exists a configuration with finite energy for which entropy is not bounded from above, or equivalently, matter can always be redistributed in such a way keeping energy fixed and increasing the entropy.

\section{Poisson equation with $\Lambda$}\label{app:B}
Let us calculate the equation of the gravitational potential $\phi$ in the presence of a cosmological constant $\Lambda$, in the Newtonian limit. The Einstein's equations are
\begin{equation}\label{eq:app1}
	R^\mu_\nu - \frac{1}{2}R\delta^\mu_\nu - \Lambda\delta^\mu_\nu = \frac{8\pi G}{c^4}T^\mu_\nu
\end{equation}
In the non-relativistic limit, the energy-momentum tensor is
\[
	T^\mu_\nu \simeq \rho c^2 \delta^\mu_0 \delta^0_\nu
\]
Contracting the Einstein's equations we get $R = -8\pi G T - 4\Lambda$ and substituting again in equation (\ref{eq:app1}) we get
\begin{equation}\label{eq:app2}
	R^\mu_\nu = \frac{4\pi G \rho}{c^2}\delta^\mu_\nu - \Lambda \delta^\mu_\nu
\end{equation}
In the weak field limit only the time components of the Einstein's equation survive. For slowly moving particles it is
\[
	\frac{d^2 x^i}{dt^2} \simeq - c^2\Gamma^i_{00} \Rightarrow \Gamma^i_{00} = \frac{1}{c^2}\partial^i\phi
\]
and in the static weak field limit it is
\[
	R^0_0 = R_{00} \simeq \frac{1}{2}\partial_\sigma\partial^\sigma g_{00} = \partial_\sigma\Gamma^\sigma_{00} = \frac{1}{c^2}\nabla^2\phi
\]
Then, the time-time component of equation (\ref{eq:app2}) gives
\[
	\nabla^2\phi = 4\pi G \rho - 8\pi G \rho_\Lambda
\]
where
\[
	\rho_\Lambda = \frac{\Lambda c^2}{8\pi G}
\]

\section{Derivation of $\delta T$}\label{app:C}
Let us give a useful expression for the dynamical energy 
\begin{eqnarray}\label{eq:appC1}
	U &=& \frac{1}{2} \int{\rho\phi_N d^3\vec{r}} + \int{\rho\phi_\Lambda d^3\vec{r}}
	 = \frac{1}{2} \int{\rho(\phi-\phi_\Lambda) d^3\vec{r}} + \int{\rho\phi_\Lambda d^3\vec{r}} \nonumber \\
	 &=& \frac{1}{2} \int{\rho\phi d^3\vec{r}} + \frac{1}{2} \int{\rho\phi_\Lambda d^3\vec{r}} 
\end{eqnarray}
We assume $\rho_\Lambda$ to be fixed. Using (\ref{eq:appC1}) we have
\begin{eqnarray}\label{eq:appC2}
	\delta E &=& \delta K + \delta U = \frac{3M}{2}\delta T + (\delta\rho)\frac{\partial U}{\partial\rho} + (\delta\phi)\frac{\partial U}{\partial\phi} + \frac{1}{2}\left((\delta\rho)\frac{\partial }{\partial\rho} + (\delta\phi)\frac{\partial }{\partial\phi} \right)^2 U + \mathcal{O}(3) \nonumber \\
	&=& \frac{3M}{2}\delta T + \frac{1}{2}\int{d^3\vec{r} (\rho\delta\phi + \phi\delta\rho + \phi_\Lambda\delta\rho)} + \frac{1}{2}\frac{\partial^2U }{\partial\rho\partial\phi}\delta\rho\delta\phi  \nonumber \\
	&=& \frac{3M}{2}\delta T + \frac{1}{2}\int{d^3\vec{r} (\rho\delta\phi_N + \phi_N\delta\rho + \phi_\Lambda\delta\rho  + \phi_\Lambda\delta\rho)} + \frac{1}{2}\int{d^3\vec{r} \delta\rho\delta\phi}  \nonumber \\
	&=& \frac{3M}{2}\delta T + \frac{1}{2}\int{d^3\vec{r} (2\phi_N\delta\rho + 2\phi_\Lambda\delta\rho)} + \frac{1}{2}\int{d^3\vec{r} \delta\rho\delta\phi}  \nonumber \\
	&=& \frac{3M}{2}\delta T + \int{d^3\vec{r} \left(\phi\delta\rho + \frac{1}{2} \delta\rho\delta\phi\right)}
\end{eqnarray}
In the third raw we used the identity
\begin{eqnarray}
	\int{d^3\vec{r} (\phi_N\delta\rho + \rho\delta\phi_N)} &=& \int{\left( d^3\vec{r}\delta\rho(r)\int{d^3\vec{r}\,'\frac{\rho(r')}{|\vec{r} - \vec{r}\,'|} }\right)} + 
	\int{\left( d^3\vec{r}\rho(r)\int{d^3\vec{r}\,'\frac{\delta\rho(r')}{|\vec{r} - \vec{r}\,'|} }\right)} \nonumber \\
	&=& \int{\left( d^3\vec{r}\delta\rho(r)\int{d^3\vec{r}\,'\frac{\rho(r')}{|\vec{r} - \vec{r}\,'|} }\right)} + 
	\int{\left( d^3\vec{r}\,'\int{d^3\vec{r}\frac{\rho(r')}{|\vec{r} - \vec{r}\,'|}\delta\rho(r) }\right)} \nonumber \\
	&=& 2\int{ d^3\vec{r}\int{d^3\vec{r}\,'\delta\rho(r)\frac{\rho(r')}{|\vec{r} - \vec{r}\,'|} }} 
	 = \int{d^3\vec{r}\, 2 \phi_N\delta\rho}
\end{eqnarray}
The constraint $\delta E = 0$ gives by use of (\ref{eq:appC2}):
\begin{equation}\label{eq:appD3}
	\delta T = - \frac{2}{3M}\int{d^3\vec{r} \left(\phi\delta\rho + \frac{1}{2} \delta\rho\delta\phi\right)}
\end{equation}

\section{How to determine the unstable branch near a turning point}\label{app:D}
We search for a solution of the problem (\ref{eq:eigen2}) for $\xi = 0$. Let $\hat{T}$ be the operator
\[
	\hat{T} = \frac{d}{dr}\left(\frac{1}{4\pi\rho r^2}\frac{d}{dr} \right) + \frac{G}{Tr^2}
\]
Let $V_T$ be the trial value we use to solve (\ref{eq:eigen2}). If it does not correspond to the solution $F_0$, it would correspond to some other solution $F_n$ for a different eigenvalue
\begin{equation}\label{eq:appD1}
	\hat{T} F_n = \frac{2\phi'}{3MT^2}V_T + \xi_n F_n
\end{equation}
so that
\[
	V_T = \int_0^Rdr\, \phi' F_n
\]
Let $\tilde{V}$ be the value of the integral
\[
	\tilde{V} = \int_0^Rdr\, \phi' F_0
\]
where $F_0$ is the solution of the problem
\begin{equation}\label{eq:appD2}
	\hat{T} F_0 = \frac{2\phi'}{3MT^2}V_T 
\end{equation}
$F_0$ will correspond to a solution with a zero eigenvalue only if
\[
	\tilde{V} = V_T
\]
Equation (\ref{eq:appD2}) gives
\[
	\int_0^Rdr\, F_0 \hat{T} F_0 = \frac{2V_T}{3MT^2}\int_0^Rdr\, \phi' F_0 = \frac{2}{3MT^2} V_T\tilde{V}
\]
and
\[
	\int_0^Rdr\, F_n \hat{T} F_0 = \frac{2V_T}{3MT^2}\int_0^Rdr\, \phi' F_n = \frac{2}{3MT^2} V_T^2
\]
Let us prove that 
\begin{equation}\label{eq:appD3}
	\int_0^R dr\, F_n \hat{T} F_0 = \int_0^R dr\, F_0 \hat{T} F_n
\end{equation}
We have
\begin{eqnarray}
	\int_0^Rdr\, F_n\hat{T} F_0 &=& \int_0^Rdr\, F_n \left\lbrace \frac{d}{dr}\left(\frac{1}{4\pi\rho r^2}\frac{dF_0}{dr} \right) + \frac{G}{Tr^2}F_0 \right\rbrace \nonumber \\
	&=&	\frac{1}{4\pi \rho r^2} \left. F_n \frac{dF_0}{dr}\right|_0^R - 
	\int_0^Rdr\, \left\lbrace \frac{dF_n}{dr}\left(\frac{1}{4\pi\rho r^2}\frac{dF_0}{dr} \right) + \frac{G}{Tr^2}F_nF_0 \right\rbrace \nonumber \\
	&=& - \frac{1}{4\pi \rho r^2} \left. \frac{dF_n}{dr}F_0\right|_0^R + \int_0^Rdr\, F_0\hat{T} F_n = \int_0^Rdr\, F_0\hat{T} F_n \nonumber
\end{eqnarray}
Using equation (\ref{eq:appD3}) and subtracting (\ref{eq:appD1}) from (\ref{eq:appD2}) we get
\[
 \hat{T} F_0 - \hat{T} F_n = -\xi_n F_n \Rightarrow \frac{2V_T^2}{3MT^2}\left(\frac{\tilde{V}}{V_T} - 1 \right)	
	= - \xi_n\int_0^R dr\, F_0F_n 
\]
Near the turning point it is $\int_0^R dr\, F_0F_n \simeq \int_0^R dr\, F_0^2 $ so that an instability ($\xi_n > 0$) sets in when $\tilde{V} < V_T$.

\section{Calculation of the total energy}\label{app:E}
	Let us calculate the expression for the total energy with no use of the virial theorem. First, we need the value $\phi(0)$ at the origin. Equation (\ref{eq:phiN}) gives
\[
	\phi(0) = -G\int_0^R dr\, \frac{\rho}{r} 4\pi r^2 = -G\int_0^z \frac{dx}{\sqrt{4\pi G \rho_0\beta}}\, \rho_0 e^{-y}4\pi\frac{x}{\sqrt{4\pi G \rho_0\beta}} \Rightarrow \beta\phi_0 = -\int_0^zdx\, x e^{-y} 
\]
In the followings we use equation (\ref{eq:appC1}). We have
\begin{eqnarray}\label{eq:appE1}
	E &=& K + U = \frac{3M}{2\beta} + \frac{1}{2} \int{\rho\phi d^3\vec{r}} + \frac{1}{2} \int{\rho\phi_\Lambda d^3\vec{r}} \Rightarrow
	\nonumber \\
	\frac{ER}{GM^2} &=& \frac{3}{2B} + \frac{R}{GM^2} \frac{1}{2} \int{\frac{1}{\beta}\rho(y+\beta\phi(0)) d^3\vec{r}} - \frac{R}{GM^2}\frac{1}{2} \frac{4\pi G}{3}\rho_\Lambda \int_0^R{\rho r^2 4\pi r^2 dr} 
	\nonumber \\
	&=& \frac{3}{2B} + \frac{1}{(\frac{GM\beta}{R})^2 R\sqrt{4\pi G \rho_0\beta}}\frac{1}{2}\int_0^ze^{-y}yx^2 +
	\frac{\beta\phi_0}{2\frac{GM\beta}{R}} -  \frac{ \frac{R}{\sqrt{4\pi G \rho_0\beta}}}{12(GM\beta)^2}\frac{2\rho_\Lambda}{\rho_0}\int_0^z x^4 e^{-y} dx \nonumber \\
	&=& \frac{3}{2B} + \frac{1}{2B^2 z}\int_0^z dx\, x^2 e^{-y} (y-\frac{\lambda}{6} x^2) -
	\frac{1}{2B} \int_0^zdx\, x e^{-y} 
\end{eqnarray}

\section*{References}

\end{document}